\documentclass[twoside,twocolumn,9pt]{article}
\usepackage{extsizes}
\usepackage[super,sort&compress,comma]{natbib} 
\usepackage[version=3]{mhchem}
\usepackage[left=1.5cm, right=1.5cm, top=1.785cm, bottom=2.0cm]{geometry}
\usepackage{balance}
\usepackage{mathptmx}
\usepackage{sectsty}
\usepackage{graphicx} 
\usepackage{lastpage}
\usepackage[format=plain,justification=justified,singlelinecheck=false,font={stretch=1.125,small,sf},labelfont=bf,labelsep=space]{caption}
\usepackage{float}
\usepackage{fancyhdr}
\usepackage{fnpos}
\usepackage[english]{babel}
\addto{\captionsenglish}{%
  
}
\usepackage{array}
\usepackage{droidsans}
\usepackage{charter}
\usepackage[T1]{fontenc}
\usepackage[usenames,dvipsnames]{xcolor}
\usepackage{setspace}
\usepackage[compact]{titlesec}
\usepackage{hyperref}
\usepackage{bm}%
\usepackage{color}
\usepackage{ulem}

\usepackage{epstopdf}

\definecolor{cream}{RGB}{222,217,201}

\begin{document}

\pagestyle{fancy}
\thispagestyle{plain}
\fancypagestyle{plain}{
\renewcommand{\headrulewidth}{0pt}
}

\makeFNbottom
\makeatletter
\renewcommand\LARGE{\@setfontsize\LARGE{15pt}{17}}
\renewcommand\Large{\@setfontsize\Large{12pt}{14}}
\renewcommand\large{\@setfontsize\large{10pt}{12}}
\renewcommand\footnotesize{\@setfontsize\footnotesize{7pt}{10}}
\makeatother

\renewcommand{\thefootnote}{\fnsymbol{footnote}}
\renewcommand\footnoterule{\vspace*{1pt}%
\color{cream}\hrule width 3.5in height 0.4pt \color{black}\vspace*{5pt}} 
\setcounter{secnumdepth}{5}

\makeatletter 
\renewcommand\@biblabel[1]{#1}            
\renewcommand\@makefntext[1]%
{\noindent\makebox[0pt][r]{\@thefnmark\,}#1}
\makeatother 
\renewcommand{\figurename}{\small{Fig.}~}
\sectionfont{\sffamily\Large}
\subsectionfont{\normalsize}
\subsubsectionfont{\bf}
\setstretch{1.125} 
\setlength{\skip\footins}{0.8cm}
\setlength{\footnotesep}{0.25cm}
\setlength{\jot}{10pt}
\titlespacing*{\section}{0pt}{4pt}{4pt}
\titlespacing*{\subsection}{0pt}{15pt}{1pt}

\fancyfoot{}
\fancyfoot[LO,RE]{\vspace{-7.1pt}\includegraphics[height=9pt]{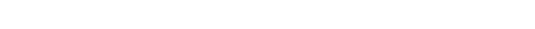}}
\fancyfoot[CO]{\vspace{-7.1pt}\hspace{13.2cm}\includegraphics{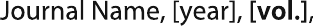}}
\fancyfoot[CE]{\vspace{-7.2pt}\hspace{-14.2cm}\includegraphics{head_foot/RF}}
\fancyfoot[RO]{\footnotesize{\sffamily{1--\pageref{LastPage} ~\textbar  \hspace{2pt}\thepage}}}
\fancyfoot[LE]{\footnotesize{\sffamily{\thepage~\textbar\hspace{3.45cm} 1--\pageref{LastPage}}}}
\fancyhead{}
\renewcommand{\headrulewidth}{0pt} 
\renewcommand{\footrulewidth}{0pt}
\setlength{\arrayrulewidth}{1pt}
\setlength{\columnsep}{6.5mm}
\setlength\bibsep{1pt}

\makeatletter 
\newlength{\figrulesep} 
\setlength{\figrulesep}{0.5\textfloatsep} 

\newcommand{\topfigrule}{\vspace*{-1pt}%
\noindent{\color{cream}\rule[-\figrulesep]{\columnwidth}{1.5pt}} }

\newcommand{\botfigrule}{\vspace*{-2pt}%
\noindent{\color{cream}\rule[\figrulesep]{\columnwidth}{1.5pt}} }

\newcommand{\dblfigrule}{\vspace*{-1pt}%
\noindent{\color{cream}\rule[-\figrulesep]{\textwidth}{1.5pt}} }

\makeatother

\twocolumn[
  \begin{@twocolumnfalse}
{\includegraphics[height=30pt]{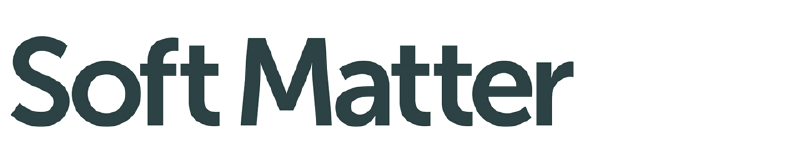}\hfill\raisebox{0pt}[0pt][0pt]{\includegraphics[height=55pt]{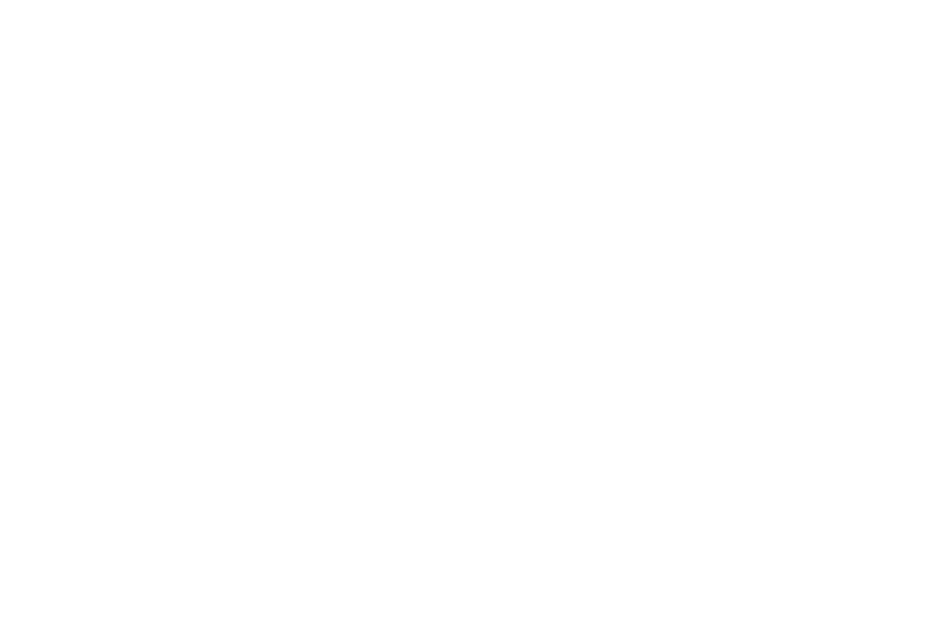}}\\[1ex]
\includegraphics[width=18.5cm]{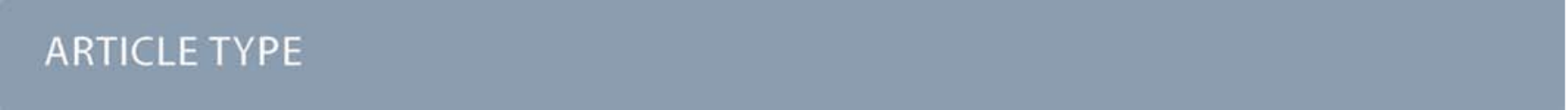}}\par
\vspace{1em}
\sffamily
\begin{tabular}{m{4.5cm} p{13.5cm} }

\includegraphics{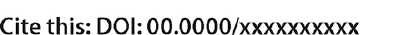} & \noindent\LARGE{\textbf{An exact expression of three-body system for the complex shear modulus of frictional granular materials}} \\
\vspace{0.3cm} & \vspace{0.3cm} \\

 & \noindent\large{Michio Otsuki\textit{$^{a}$} and Hisao Hayakawa\textit{$^{b}$}} \\

\includegraphics{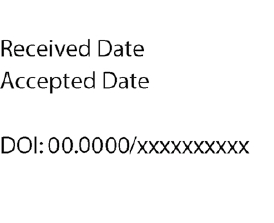} & \noindent\normalsize{
We propose a simple model comprising three particles to study the nonlinear mechanical response of jammed frictional granular materials under oscillatory shear.
Owing to the introduction of the simple model, we obtain an exact analytical expression of the complex shear modulus for a system including many monodispersed disks, which satisfies a scaling law in the vicinity of the jamming point.
These expressions perfectly reproduce the shear modulus of the many-body system with low strain amplitudes and friction coefficients.
Even for disordered many-body systems, the model reproduces results by introducing a single fitting parameter.
} \\

\end{tabular}

 \end{@twocolumnfalse} \vspace{0.6cm}

  ]

\renewcommand*\rmdefault{bch}\normalfont\upshape
\rmfamily
\section*{}
\vspace{-1cm}


\footnotetext{\textit{$^{a}$~  Graduate School of Engineering Science, Osaka University, Toyonaka, Osaka 560-8531, Japan. E-mail: m.otsuki.es@osaka-u.ac.jp}}
\footnotetext{\textit{$^{b}$~Yukawa Institute for Theoretical Physics, Kyoto University, Kitashirakawaoiwake-cho, Sakyo-ku, Kyoto 606-8502, Japan}. }





\section{Introduction}
The rheological property of densely dispersed grains, e.g., granular materials, colloidal suspensions, and emulsions, plays an important role in physics and engineering. 
This rheological property mainly depends on the packing fraction $\phi$ of the grains.
The materials behave like fluids for $\phi<\phi_{\rm J}$ with jamming fraction $\phi_{\rm J}$ and exhibit a solid-like elastic response above $\phi_{\rm J} $\cite{Hecke,Behringer}.
In the linear response regime (i.e., for small strains), the shear modulus is characterized by the density of states \cite{Maloney2004,Maloney2006,Ishima22} and satisfies scaling laws \cite{OHern02,Wyart,Tighe11,Otsuki17}.
However, the linear response region becomes narrower as $\phi$ approaches $\phi_{\rm J}$ \cite{Coulais,Otsuki14}, and the nonlinear response becomes relevant 
due to the plastic deformation associated with the yielding \cite{Nagamanasa,Knowlton,Kawasaki16,Leishangthem,Clark,Boschan19,Boschan,Nakayama,Kawasaki20}.

If we are interested in a nonlinear response to an applied oscillatory shear strain, it exhibits a complicated stress-strain curve. 
Although the storage and loss moduli $G'$ and $G''$ were originally introduced to characterize the linear viscoelasticity of materials, they can use to characterize nonlinear viscoelasticity or visco-elastoplastic responses to applied strains \cite{Hyun2011}.
In this case, G' and G" are no longer constants but strongly depend on the strain amplitude $\gamma_0$.
In particular, we have recognized that $G'$ decreases with $\gamma_0$ \cite{Otsuki14,Bohy,Ishima,Otsuki21,Otsuki22} and $G"$ remains non-zero in the low frequency limit\cite{Otsuki21,Otsuki22} for densely dispersed grains. 

The theoretical analysis of densely dispersed grains is challenging as a typical many-body problem in non-equilibrium systems.
To date, a few theoretical approaches have been proposed for systems related to frictionless particles.
The scaling laws for the linear elastic response were derived in terms of the vibrational density of states \cite{Wyart,Tighe11}.
The Fourier analysis of particle trajectories helps to generate semi-analytical expressions for $G'$ and $G''$ \cite{Otsuki22}.
Unfortunately, these theories can not apply to frictional particles because of the history-dependent contact force \cite{Otsuki17,Otsuki21}.

It is helpful to analyze a simple model with small degrees of freedom to understand the behavior of many-body systems, including densely dispersed grains.
This approach has been used in statistical mechanics.
The mean-field approximation of the Ising model is a typical example in which the system contains only one Ising spin under the influence of a self-consistently determined mean field \cite{Goldenfeld}.
For atomic liquids, a cell model, in which a single atom exists in a cage, was used to derive the equation of state \cite{Lennard37,Lennard38}. 
The coherent potential approximation for disordered solids has been used to understand electronic band structures \cite{Yonezawa}.
The effective medium theory reveals the elastic response of random spring networks \cite{Feng}.
In addition, a simple model consisting of two particles was proposed to reproduce the liquid-solid phase transition \cite{Awazu}.
The advantage of such few-body models is that we can obtain exact solutions. 
The qualitative behavior of the corresponding many-body systems can be determined based on the solutions of the few-body models.
Thus, we adopt this approach to determine the nonlinear responses of the frictional dispersed grains.

This study proposes a model consisting of three identical particles to describe the mechanical response of jammed frictional granular materials under oscillatory shear.
In Section \ref{Sec:TBS}, we introduce the three-body system (TBS).
This model can be analytically solved for low-strain amplitudes and friction coefficients near the jamming point in Section \ref{Sec:Theory}.
In Section \ref{Sec:MBS}, we demonstrate that the analytical solution reproduces the storage and loss moduli of many-body systems (MBSs) without any fitting parameter if there is no disorder in the particle configuration.
Even if disorder exists, a scaling law for the complex shear modulus for the TBS semi-quantitatively agrees with the numerical simulations of the MBS by introducing a fitting parameter.
We discuss and conclude our results in Section \ref{Sec:Conclusion}.
In Appendix A, we show the details of the MBS when the particles are initially placed on a triangular lattice.
The effect of particle rotation is described in Appendix B.
In Appendix C, we derive the analytical expressions for the shear stress and pressure in the TBS.
In Appendix D, we relate the complex shear modulus with the hysteresis loop of the stress--strain curve.
The details of the disordered MBS are presented in Appendix E.
We present the numerical shear modulus for the TBS in Appendix F.

\section{Three-Body System}
\label{Sec:TBS}

\begin{figure}[htbp]
\centering
\includegraphics[width=0.8\linewidth]{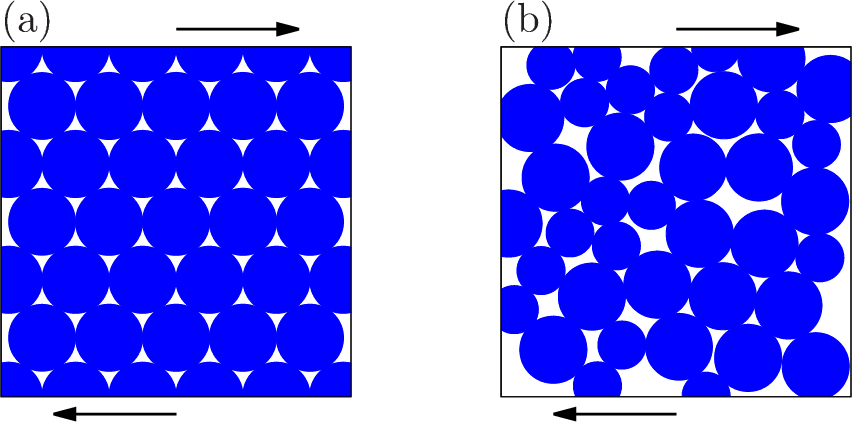}
  \caption{
    Schematics of the ordered MBS (a) and the disordered MBS (b).
}
\label{confMBS}
\end{figure}

\begin{figure}[htbp]
\centering
\includegraphics[width=0.6\linewidth]{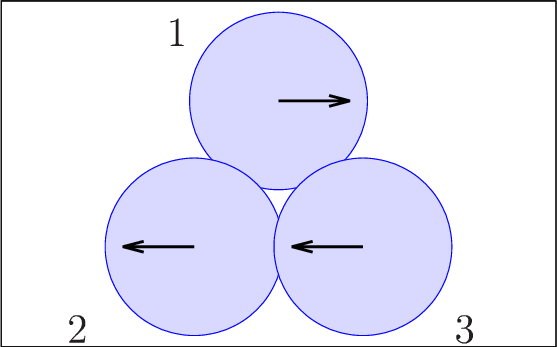}
  \caption{
    A schematic of the TBS.
}
\label{Fig1}
\end{figure}

We consider two-dimensional granular materials consisting of many grains under oscillatory shear (Fig. \ref{confMBS}).
Here, the grains constituting granular materials are modeled as frictional spherical particles.
Moreover, we introduce a system of three identical particles to simply describe the MBS (Fig. \ref{Fig1}).
The MBS can contain polydisperse particles, while we assume that the TBS is a monodisperse system.
In the TBS, the position $\boldsymbol r_i(t) = (x_i(t),y_i(t))$ of particle $i$ with diameter $d$ at time $t$ is given by
\begin{eqnarray}
\label{TE1}
  \boldsymbol r_1(t) & = & \left ( \frac{\sqrt{3} \gamma(\theta(t)) \ell}{4}, \frac{\sqrt{3} \ell} {4} \right), \\
\label{TE2}
  \boldsymbol r_{2}(t) & = & \left ( - \frac{\sqrt{3} \gamma(\theta(t)) \ell }{4} - \frac{ \ell}{ 2}, - \frac{\sqrt{3} \ell}{4} \right), \\
\label{TE3}
  \boldsymbol r_{3}(t) & = & \left ( - \frac{\sqrt{3} \gamma(\theta(t)) \ell }{4} + \frac{ \ell}{ 2}, - \frac{\sqrt{3} \ell}{4} \right),
\end{eqnarray}
where $\ell$ is the initial distance between particles.
We also introduce $\varepsilon:=1-\ell/d$ as the compressive strain.
The compressive strain $\epsilon$ in the TBS corresponds to $\phi-\phi_{\rm J}$ in the MBS as shown in Appendix A.
We apply shear strain as
\begin{equation}
  \gamma(\theta) = \gamma_0 \sin  \theta
  \label{g}
\end{equation}
with strain amplitude $\gamma_0$, phase $\theta = \omega t$, and angular frequency $\omega$.
Note that we need at least three particles to realize a stable interlocking state.

We adopt the interaction force $\boldsymbol f_{ij}$ between particles $i$ and $j$ given by
\begin{equation}
  \label{F}
  \boldsymbol f_{ij} = \left ( f_{ij}^{\rm (n)} \boldsymbol n_{ij}    + f_{ij}^{\rm (t)}\boldsymbol t_{ij}  \right )H(r_{ij}-d),
\end{equation}
where $f_{ij}^{\rm (n)}$ and $f_{ij}^{\rm (t) }$ denote the normal and tangential forces between the particles $i$ and $j$ \cite{DEM}.
The distance between the particles $i$ and $j$ is $r_{ij} = |\boldsymbol r_{ij}|$ with $\boldsymbol r_{ij} := \boldsymbol r_i - \boldsymbol r_j = (x_{ij},y_{ij})$.
Here, $H(x)$ is Heviside's step function satisfying $H(x)=1$ for $x>0$ and $H(x) = 0$ otherwise.
The normal and tangential unit vectors are denoted by $\boldsymbol n_{ij} := \boldsymbol r_{ij}/r_{ij} = (n_{ij,x}, n_{ij,y})$ and $\boldsymbol t_{ij} := (-n_{ij,y},n_{ij,x})$, respectively.
For simplicity, we do not consider the torque balance and, thus, the rotation of the particles.
See Appendix B for the effect of the rotation.

The normal force is assumed to be
\begin{equation}
f_{ij}^{\rm (n)} = - k_{\rm n} u^{\rm (n)}_{ij} 
\label{Fn}
\end{equation}
with the normal elastic constant $k_{\rm n}$ and normal relative displacement $u^{\rm (n)}_{ij} := r_{ij} - d$.
Moreover, the tangential force is assumed to be
\begin{equation}
f_{ij}^{\rm (t)} = {\rm min} \left ( |\tilde f_{ij}^{\rm (t)}|, \mu f_{ij}^{\rm (n)} \right ) {\rm sgn} (\tilde f_{ij}^{\rm (t)}),
  \label{Ft}
\end{equation}
where  $\tilde f_{ij}^{\rm (t)} = - k_{\rm t} u_{ij}^{\rm (t)}$; $k_{\rm t}$ denotes the tangential elastic constant, and $\mu$ denotes the friction coefficient.
Here, ${\rm min}(a,b)$ selects the smaller value between $a$ and $b$, ${\rm sgn}(x) = 1$ for $x \ge 0$, and ${\rm sgn}(x) = -1$ for $x < 0$. 
The tangential displacement $u_{ij}^{\rm (t)}$ satisfies $\frac{d}{dt} u_{ij}^{\rm (t)} = v_{ij}^{\rm (t)}$ for 
$|\tilde f_{ij}^{\rm (t)}| <\mu f_{ij}^{\rm (n)}$
with the tangential velocity $v_{ij}^{\rm (t)} = (\frac{d}{dt} {\boldsymbol  r}_i - \frac{d}{dt}{ \boldsymbol r}_j)\cdot \boldsymbol t_{ij}$, whereas $u_{ij}^{\rm (t)}$ remains unchanged for
$|\tilde f_{ij}^{\rm (t)}| \ge \mu f_{ij}^{\rm (n)}$.
We refer to the contact with $|\tilde f_{ij}^{\rm (t)}| < \mu f_{ij}^{\rm (n)}$ as the stick contact and the contact with $|\tilde f_{ij}^{\rm (t)}| \ge \mu f_{ij}^{\rm (n)}$ as the slip contact.
The tangential displacement, $u_{ij}^{\rm (t)}$, is initially set to zero.

The (symmetric contact) shear stress is given by
\begin{equation}
  \sigma(\theta;\gamma_0,\mu)  = \sigma^{\rm (n)} (\theta;\gamma_0,\mu)+\sigma^{\rm (t)} (\theta;\gamma_0,\mu)
  \label{s}
\end{equation}
with the normal component of $\sigma$
\begin{equation}
  \sigma^{\rm (n)}(\theta;\gamma_0,\mu) = - \frac{1}{A} \sum _i \sum_{j>i} \frac{x_{ij}y_{ij}}{r_{ij}} f^{\rm (n)}_{ij}
  \label{Sn}
\end{equation}
and tangential component of $\sigma$
\begin{equation}
  \sigma^{\rm (t)}(\theta;\gamma_0,\mu) = - \frac{1}{2A} \sum _i \sum_{j>i} \frac{x_{ij}^2-y_{ij}^2}{r_{ij}} f^{\rm (t)}_{ij}.
  \label{St}
\end{equation}
Here, $A$ corresponds to the area of the system, and we choose $A = \sqrt{3} \ell^2 /2$ as shown in Appendix A.
The pressure is given by
\begin{equation}
  P(\theta;\gamma_0,\mu) =  \frac{1}{2A} \sum _i \sum_{j>i} (x_{ij} f_{ij,x}+y_{ij} f_{ij,y}).
  \label{P}
\end{equation}
In the right-hand sides of eqns. \eqref{Sn}-\eqref{P}, we have omitted the arguments $\theta$, $\gamma_0$, and $\mu$. 
Similar abbreviations are used below.
As we are interested in quasistatic processes, we do not consider the kinetic parts of $\sigma$ and $P$ and the dependence on $\omega$.
After several cycles of oscillatory shear, $\sigma(\theta)$ becomes periodic.
The storage and loss moduli are given by \cite{Doi}
\begin{eqnarray}
  G'(\gamma_0,\mu) & = & \frac{1}{\pi} \int_0^{2\pi} d \theta \ \sigma(\theta;\gamma_0,\mu) \sin \theta / \gamma_0, \label{Gp}\\
  G''(\gamma_0,\mu) & = & \frac{1}{\pi} \int_0^{2\pi} d \theta \ \sigma(\theta;\gamma_0,\mu) \cos \theta / \gamma_0 \label{Gpp}.
\end{eqnarray}

\section{Theoretical analysis} 
\label{Sec:Theory}

Assuming $\gamma_0 \ll \epsilon \ll 1$, we analytically obtain $G'$ and $G''$ for the TBS.
The derivation of the analytical results can be found in Appendix C.

First, the normal component of the shear stress is given by
\begin{eqnarray}
  \label{sn}
  \sigma^{\rm (n)}(\theta) & = & \frac{\sqrt{3}k_{\rm n} \gamma(\theta)}{4}.
\end{eqnarray}
The tangential component of the shear stress is given by
\begin{eqnarray}
  \label{st1}
  \sigma^{\rm (t)}(\theta) & = & \frac{\sqrt{3}k_{\rm t} \gamma(\theta)}{4}
\end{eqnarray}
for $\gamma_0 < \gamma_c(\mu)$ with a critical amplitude
\begin{equation}
  \gamma_c(\mu) = \frac{4 \mu  k_{\rm n} \epsilon}{ 3 k_{\rm t}},
  \label{gc}
\end{equation}
which characterizes the transition from stick to slip states in the contact between the particles.
For $\gamma_0 \ge \gamma_c(\mu)$, the tangential component of the shear stress is given by
\begin{eqnarray}
  \label{st2}
  &\sigma^{\rm (t)}(\theta)  =  
   \left\{ 
\begin{array}{ll}
  \dfrac{\mu k_{\rm n} \epsilon}{\sqrt{3}} , & 0 \le \theta < \dfrac{\pi}{2} \\
  \dfrac{\mu k_{\rm n} \epsilon}{\sqrt{3}}  + \dfrac{\sqrt{3}k_{\rm t} (\gamma(\theta) - \gamma_0)}{4}, & \dfrac{\pi}{2} \le \theta < \dfrac{\pi}{2} + \Theta \\
  -\dfrac{\mu k_{\rm n} \epsilon}{\sqrt{3}} , & \dfrac{\pi}{2} + \Theta \le \theta <  \dfrac{3\pi}{2}\\
  -\dfrac{\mu k_{\rm n} \epsilon}{\sqrt{3}} + \dfrac{\sqrt{3}k_{\rm t} (\gamma(\theta) + \gamma_0)}{4}, & \dfrac{3\pi}{2} \le \theta <  \dfrac{3\pi}{2} + \Theta\\
  \dfrac{\mu k_{\rm n} \epsilon}{\sqrt{3}} , & \dfrac{3\pi}{2} + \Theta \le \theta <  2\pi,
\end{array}
\right.
\end{eqnarray}
where $\Theta = \cos^{-1} \left ( 1 - 2 \gamma_c(\mu) / \gamma_0 \right )$.
Regions with $\frac{\pi}{2} \le \theta < \frac{\pi}{2} + \Theta$ and $\frac{3\pi}{2} \le \theta <  \frac{3\pi}{2} + \Theta$ correspond to the stick state, and the other regions correspond to the slip state.
Owing to this transition in the contact, the stress--strain curve given by eqns. \eqref{sn}--\eqref{st2} exhibits a hysteresis loop.
Equation \eqref{st2} does not exhibit a viscoelastic response but a typical elastoplastic response without viscous effect.

\begin{figure}[htbp]
\centering
\includegraphics[width=1.0\linewidth]{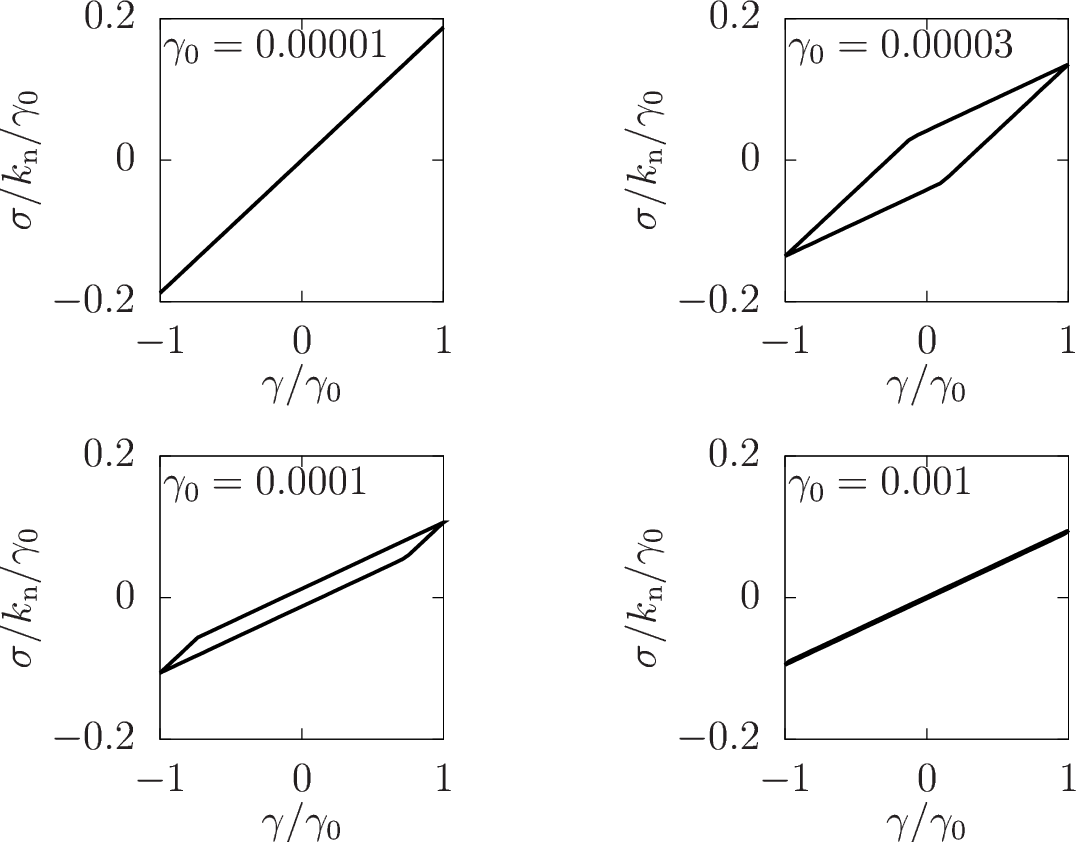}
  \caption{
    Scaled shear stress $\sigma/\gamma_0$ against $\gamma/\gamma_0$ 
    using eqns. \eqref{g}, \eqref{s}, and \eqref{sn}-\eqref{st2} 
    for various values of $\gamma_0$ with $k_{\rm t}/k_{\rm n}=1.0$, $\epsilon = 0.001$, and $\mu = 0.01$. 
}
\label{st:Fig}
\end{figure}

Figure \ref{st:Fig} shows the scaled shear stress $\sigma/\gamma_0$ against the scaled strain $\gamma/\gamma_0$ using eqns. \eqref{g}, \eqref{s}, and \eqref{sn}-\eqref{st2} for various values of $\gamma_0$ with $k_{\rm t}/k_{\rm n}=1.0$ and $\mu = 0.01$.
The shape of the scaled stress--strain curve is characterized by a parallelogram as a typical elastoplastic response.
As $\gamma_0$ increases, the maximum value $\tilde \sigma_{\rm max} =( \sigma / \gamma_0)|_{\gamma/\gamma_0 = 1}$ decreases from a larger value $\sqrt{3}(k_{\rm n}+k_{\rm t})/4$ to a smaller value $\sqrt{3}k_{\rm n}/4$.
As shown in Appendix D, the storage modulus $G'$ is approximately given by $\tilde \sigma_{\rm max}$.
Hence, the decrease of $\tilde \sigma_{\rm max}$ in Fig. \ref{st:Fig} indicates the decrease of $G'$.
For $\gamma_0 = 0.00003$ and $0.0001$, the hysteresis loop exists, but the area of the loop is negligible for $\gamma_0 = 0.00001$ and $0.001$.
The loss modulus $G''$ is proportional to the area of the loop as shown in Appendix D.
Hence, the dependence of the area on $\gamma_0$ indicates that there is a peak in $G''$ as $\gamma_0$ increases.

Substituting eqns. \eqref{s} and \eqref{sn}-\eqref{st2} into eqn. \eqref{Gp}, we obtain the storage modulus as
\begin{align}
  \label{GpT}
  & G'  =  
   \left\{
\begin{array}{ll}
  \dfrac{\sqrt{3}\left( k_{\rm n} + k_{\rm t}\right )}{ 4}, & \gamma_0 \le \gamma_c(\mu) \\
  \dfrac{\sqrt{3}}{ 4}\left\{ k_{\rm n} + \dfrac{k_{\rm t}}{\pi} \left ( \Theta - \sin \Theta \cos \Theta \right ) \right \}, & \gamma_0 > \gamma_c(\mu).
\end{array}
\right. 
\end{align}
As $\gamma_0$ increases beyond $\gamma_c(\mu)$, $G'$ decreases from a higher value to a lower value.
The corresponding behavior has been observed in the MBS in previous studies \cite{Otsuki17,Otsuki21}.

Substituting eqns. \eqref{s} and \eqref{sn}-\eqref{st2} into eqn. \eqref{Gpp}, the loss modulus is given by
\begin{align}
  \label{GppT}
  & G''  =  
   \left\{
\begin{array}{ll}
  0, & \gamma_0 \le \gamma_c(\mu) \\
  \dfrac{\sqrt{3}k_{\rm t}}{ 4  \pi}\left ( 1 -  \cos^2  \Theta \right ), & \gamma_0 > \gamma_c(\mu) .
\end{array}
\right.
\end{align}
The loss modulus $G''$ is zero for $\gamma_0 < \gamma_c(\mu)$, whereas $G''$ sharply increases with $\gamma_0$ when $\gamma_0$ exceeds $\gamma_c(\mu)$ and decreases to $0$ after a peak.
The behavior of $G''$ for the TBS qualitatively reproduces that of the MBS in previous studies \cite{Otsuki21}.

We adopt the abbreviation for the pressure at $\gamma=0$ as
\begin{equation}
P_0 := P(\theta=0;\gamma_0,\mu),
\end{equation}
which is also obtained as
\begin{equation}
  P_0  = \sqrt{3}k_{\rm n} \epsilon.
  \label{PT}
\end{equation}
From eqns. \eqref{gc}, \eqref{GpT}, \eqref{GppT}, and \eqref{PT}, we derive scaling laws for a given $\epsilon$ as
\begin{align}
  \label{Gp:scale}
  & G'(\mu,\gamma_0) = G'_{\rm M}(\mu) \mathcal{G}'\left( \frac{k_{\rm t}\gamma_0}{  \mu  P_0(\gamma_0,\mu)} \right ), \\
  \label{Gpp:scale}
  & G''(\mu,\gamma_0) = G''_{\rm M}(\mu) \mathcal{G}''\left( \frac{k_{\rm t} \gamma_0}{ \mu  P_0(\gamma_0,\mu)} \right ),
\end{align}
where $\mathcal{G}'(x)$ and $\mathcal{G}''(x)$ denote scaling functions.
The maximum values of $G'$ and $G''$ are denoted as $G'_{\rm M}$ and $G''_{\rm M}$, respectively.
In the TBS, they are given as
\begin{align}
& G'_{\rm M} = \sqrt{3}\left( k_{\rm n} + k_{\rm t}\right )/4, \ \ 
 G''_{\rm M} = \sqrt{3}k_{\rm t}/(4 \pi), \\
  \label{GpS}
  & \mathcal{G}'(x)  =
   \left\{
\begin{array}{ll}
  1, & x \le x_c, \\
  \left ( 1 + \dfrac{k_{\rm t}}{k_{\rm n}}\dfrac{T(x) - S(x)}{\pi}  \right )/
  \left (1+\dfrac{k_{\rm t}}{k_{\rm n}} \right ), & x > x_c,
\end{array}
\right. \\
  \label{GppS}
  & \mathcal{G}''(x)  = 
   \left\{
\begin{array}{ll}
  0, & x \le x_c, \\
  1 -  \cos^2  T(x) , & x >x_c
\end{array}
\right.
\end{align}
with $T(x) = \cos^{-1}(1-2 x_c / x)$, $S(x) = \sin (2T(x))/2$, and $x_c = 4 /(3 \sqrt{3})$.

\section{Comparison with the MBS} 
\label{Sec:MBS}

We demonstrate the relevance of the TBS analysis based on the simulation of a two-dimensional MBS consisting of $N$ frictional grains.
First, we consider a system corresponding to the TBS, where all the particles are identical and initially placed on the triangular lattice with a unit length $\ell$ (Fig. \ref{confMBS}(a)).
The details are shown in Appendix A.
Next, we consider a bidisperse system where the number of particles with diameter $d$ is equal to that of particles with diameter $d/1.4$, and the particles are randomly placed with packing fraction $\phi$ (Fig. \ref{confMBS}(b)).
The mass densities of the particles are identical.
The details of the disordered MBS are shown in Appendix E.
In both systems, the shear strain given by eqn. \eqref{g} is applied for $N_{\rm c}$ cycles using the SLLOD equation under the Lees--Edwards boundary condition \cite{Evans}.
In the MBS, we replace the normal force as
\begin{equation}
 f_{ij}^{\rm (n)} \to -\left ( k_{\rm n} u^{\rm (n)}_{ij}
 + \eta_{\rm n} v^{\rm (n)}_{ij}\right )
\end{equation}
with the normal viscous constant $\eta_{\rm n}$ and the normal velocity $v_{ij}^{\rm (n)} = (\frac{d}{dt} {\boldsymbol  r}_i - \frac{d}{dt}{ \boldsymbol r}_j)\cdot \boldsymbol n_{ij}$
to include the viscous force depending on the relative velocity.
The tangential force is replaced by
\begin{equation}
  f_{ij}^{\rm (t)} \to {\rm min} \left ( |\tilde f_{ij}^{\rm (t)}|, \mu f_{ij}^{\rm (n,el)} \right ) {\rm sgn} (\tilde f_{ij}^{\rm (t)}),
\end{equation}
with 
\begin{equation}
  \tilde f_{ij}^{\rm (t)} \to - \left ( k_{\rm t} u_{ij}^{\rm (t)} + \eta_{\rm t} v_{ij}^{\rm (t)} \right ),
\end{equation}
where $f_{ij}^{\rm (n,el)} = - k_{\rm n} u^{\rm (n)}_{ij}$ denotes the elastic part of the normal force with a tangential viscous constant $\eta_{\rm t}$.
We measure $G'$, $G''$, and $P_0$ in the last cycle using eqns. \eqref{P}-\eqref{Gpp}.
For the ordered MBS, we use $N=64$, $k_{\rm t}/k_{\rm n}=1.0$, and $\epsilon = 0.001$, whereas $N=1000$, $k_{\rm t}/k_{\rm n}=0.2$, and $\phi=0.87$ are used for the disordered MBS.
In both systems, the other parameters are identical: $N_{\rm c}=20$, $\eta_{\rm t} = \eta_{\rm n} = \sqrt{mk_{\rm n}}$, and $\omega = 0.0001 \sqrt{m/k_{\rm n}}$ with a mass $m$ of larger particles.

\begin{figure}[htbp]
\centering
\includegraphics[width=0.7\linewidth]{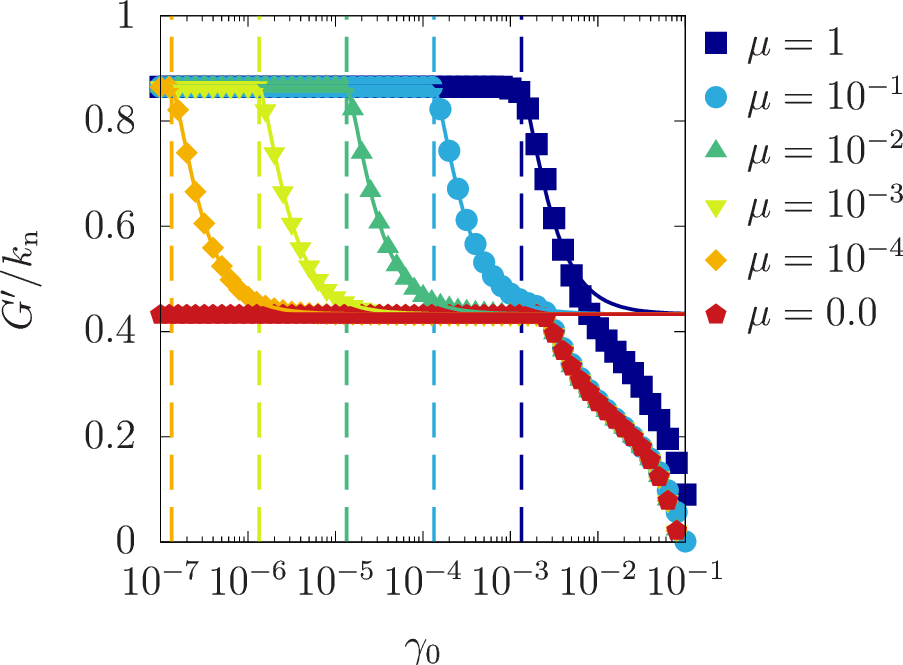}
  \caption{
    Storage modulus $G'$ against $\gamma_0$ with $k_{\rm t} / k_{\rm n} = 1.0$ and $\epsilon = 0.001$ for various values of $\mu$.
    The points represent the results of the ordered MBS.
    The thin solid lines represent the analytical result given by eqn. \eqref{GpT}. 
    The vertical dashed lines represent the critical amplitude $\gamma_c(\mu)$ given by eqn. \eqref{gc} for $\mu = 10^{-4}, 10^{-3}, 10^{-2}, 10^{-1}$, and $1$ from left to right.
}
  \label{Gp_3P}
\end{figure}

As shown in Fig. \ref{Gp_3P}, we plot $G'$ for the ordered MBS against $\gamma_0$ with $k_{\rm t}/k_{\rm n} = 1.0$ and $\epsilon = 0.001$ for various values of $\mu$ as points.
Moreover, we plot the analytical results of the TBS obtained using eqn. \eqref{GpT} as thin solid lines.
Surprisingly, the results of the TBS agree with those of the MBS for $\gamma_0 < 0.003$ without any fitting parameters.
As $\gamma_0$ increases beyond $\gamma_c(\mu)$ shown by the vertical dashed lines, $G'$ for $\mu>0$ decreases and converges to a constant, which is equal to $G'$ for $\mu=0$.
For larger $\gamma_0$, $G'$ for the MBS decreases again, whereas the theoretical $G'$ for the TBS is constant.
This discrepancy results from the violation of condition $\gamma_0 \ll \epsilon$ for the analytical calculation.
If we numerically solve the TBS to obtain $G'$ without the assumption $\gamma_0 \ll \epsilon$, $G’$ decreases after a plateau again as in the case of MBS, although its value in the TBS for $\gamma_0 \to 0.1$ slightly deviates from that of the MBS, as shown in Appendix F.

As shown in Fig. \ref{Gpp_3P}, we plot $G''$ for the MBS on the triangular lattice against $\gamma_0$ with $k_{\rm t}/k_{\rm n} = 1.0$ and $\epsilon = 0.001$ for various values of $\mu$ as points.
Moreover, we plot the analytical results of the TBS obtained using eqn. \eqref{GppT} as thin solid lines.
The analytical result agrees perfectly with the MBS for $\gamma_0 < 0.003$ without any fitting parameters.
As $\gamma_0$ increases beyond $\gamma_c(\mu)$ shown by the vertical dashed lines, $G''$ for $\mu>0$ increases from $0$ and decreases after reaching a peak.
The peak position of $G''$ against $\gamma_0$ increases with $\mu$.
Thus, our analytical results fail to capture the behavior of $G''$ for $\mu=1$.

\begin{figure}[htbp]
\centering
\includegraphics[width=0.7\linewidth]{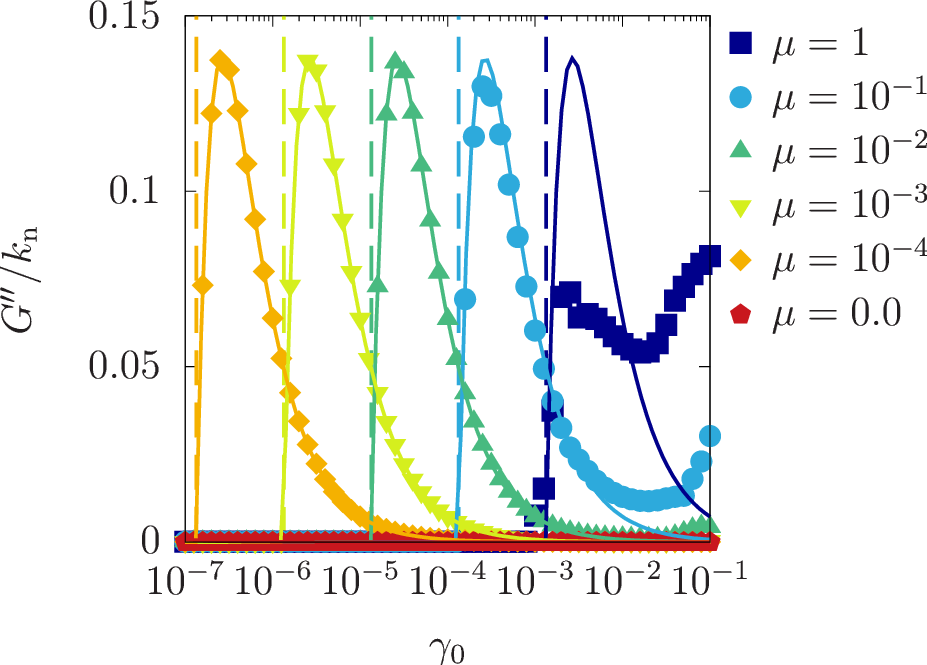}
  \caption{
    Loss modulus $G''$ against $\gamma_0$ with $k_{\rm t} / k_{\rm n} = 1.0$ and $\epsilon = 0.001$ for various values of $\mu$.
    The points represent the results of the ordered MBS.
    The thin solid lines represent the analytical results obtained using eqn. \eqref{GppT}. 
    The vertical dashed lines represent the critical amplitude $\gamma_c(\mu)$ given by eqn. \eqref{gc} for $\mu = 10^{-4}, 10^{-3}, 10^{-2}, 10^{-1}$, and $1$ from left to right.
}
  \label{Gpp_3P}
\end{figure}

\begin{figure}[htbp]
\centering
\includegraphics[width=1.0\linewidth]{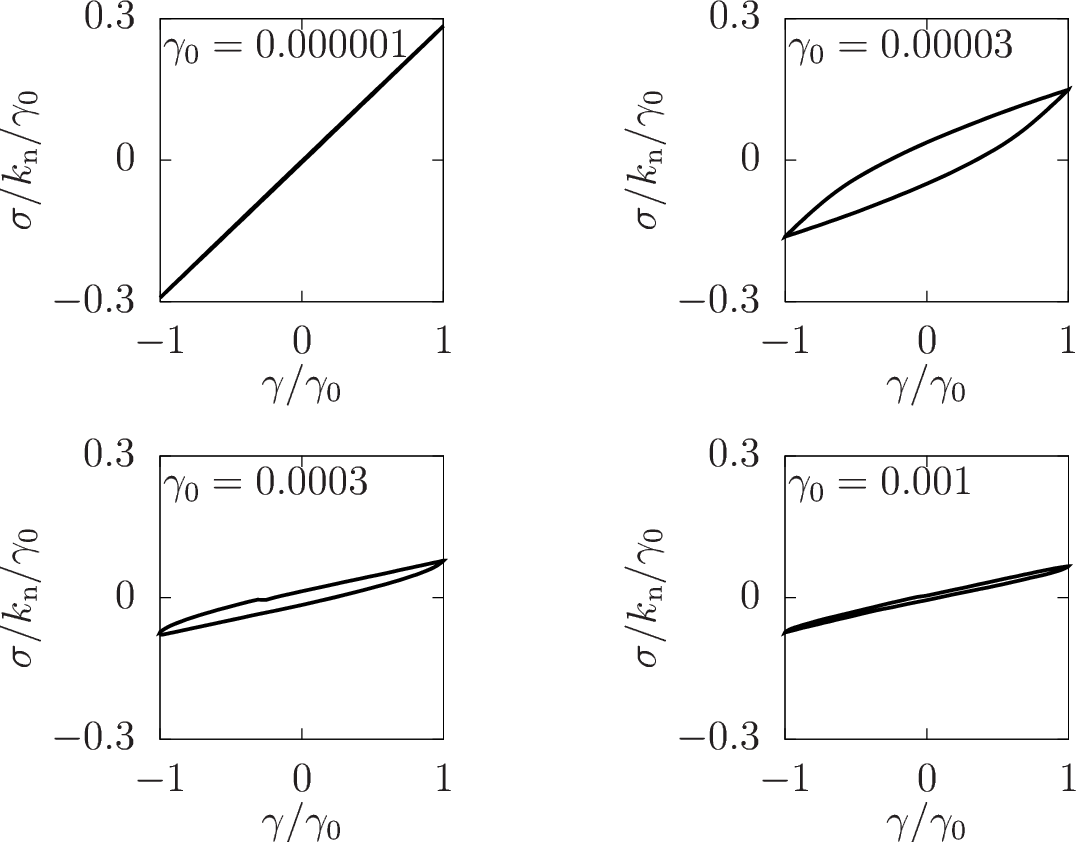}
  \caption{
    Scaled shear stress $\sigma/\gamma_0$ against $\gamma/\gamma_0$ in the disordered MBS for various values of $\gamma_0$ with $\mu = 0.01$ and $\phi=0.870$. 
}
\label{st_YPT10:Fig}
\end{figure}

Consider the disordered MBS shown in Fig. \ref{confMBS}(b).
Figure \ref{st_YPT10:Fig} shows the scaled shear stress $\sigma/\gamma_0$ against the scaled strain $\gamma/\gamma_0$ in the disordered MBS with $\mu = 0.0001$. 
The maximum value $\tilde \sigma_{\rm max}$ decreases as $\gamma_0$ increases.
The area $S$ of the curve is the largest for $\gamma_0 = 0.00003$.
It is interesting that stress-strain curves are not characterized by parallelograms in this case in contrast to Fig. \ref{st:Fig}.
This means that the disordered configuration of particles creates an effective viscosity, and thus, the response to an applied strain becomes visco-elastoplastic.

\begin{figure}[htbp]
\centering
\includegraphics[width=1.0\linewidth]{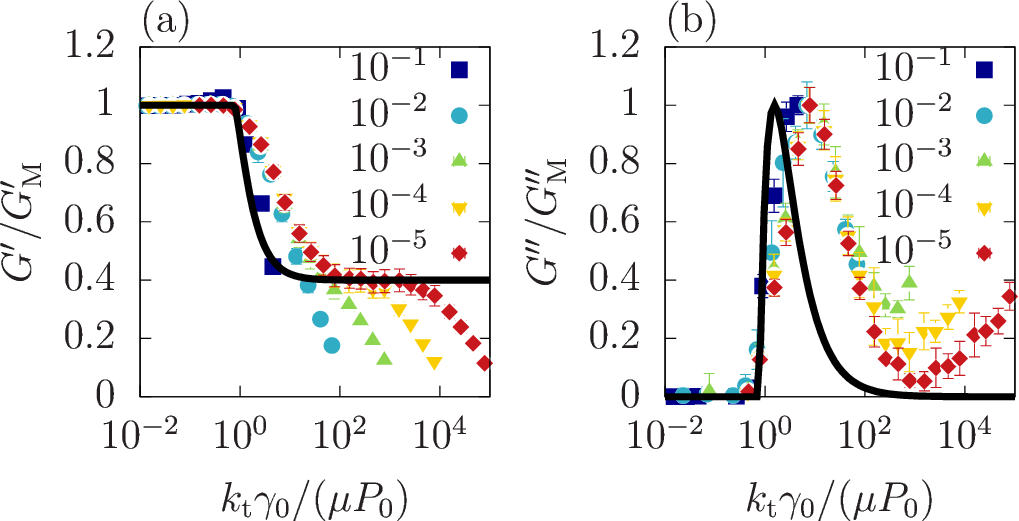}
  \caption{
(a) Scaled storage modulus $G'/G'_{\rm M}$ against the scaled strain $k_{\rm t} \gamma_0/ (\mu P_0(\gamma_0,\mu))$ with $\phi=0.87$ and $k_{\rm t}/k_{\rm n}=0.2$ for various values of $\mu$ in the disordered MBS.
The solid line represents the analytical result of the TBS given by eqn. \eqref{GpS} with $k_{\rm t}/k_{\rm n}=1.5$.
(b) Scaled loss modulus $G''/G''_{\rm M}$ against the scaled strain $k_{\rm t} \gamma_0/ (\mu P_0(\gamma_0,\mu))$ with $\phi=0.87$ and $k_{\rm t}/k_{\rm n}=0.2$ for various values of $\mu$ in the disordered MBS.
The solid line represents the analytical result of the TBS given by eqn. \eqref{GppS} with $k_{\rm t}/k_{\rm n}=1.5$.
}
  \label{YPT10_scale}
\end{figure}

The behaviors of $G'$ and $G''$ in the disordered MBS are similar to those of the TBS as shown in Appendix E.
Therefore, it is expected that the scaling laws in eqns.  \eqref{Gp:scale} and \eqref{Gpp:scale} for a given $\epsilon$ in the TBS can be used even in this system with corresponding $\phi$. 
This expectation is verified by Fig. \ref{YPT10_scale}, in which we plot the scaled moduli $G'/G'_{\rm M}$ and $G''/G''_{\rm M}$ against the scaled strain $k_{\rm t} \gamma_0/ (\mu P_0(\gamma_0,\mu))$ for various values of $\mu$ in the disordered MBS. 
Moreover, we plot the analytical results for the TBS obtained using eqns. \eqref{GpS} and \eqref{GppS} as solid lines, which qualitatively reproduce the MBS results for small scaled strain, while the scaling is apparently violated for large scaled strain.
Here, we choose $k_{\rm t}/k_{\rm n}=1.5$ for the TBS to fit the second plateau to that of the MBS.
At present, we do not know the relationship between $\phi$ and the fitting parameter.

\section{Conclusions}
\label{Sec:Conclusion}

We demonstrated the relevancy of a model of the TBS to describe the complex modulus of jammed frictional granular materials under oscillatory shear.
We obtained the analytical expressions for the $\gamma_0$-dependence of $G'$ and $G''$ as shown in eqns. \eqref{gc}, \eqref{GpT}, and \eqref{GppT}, which predict the $\mu$-dependence of the critical amplitude $\gamma_c$, the decrease of $G'$, and the peak of $G''$ above $\gamma_c$ for crystalline solids.
The analytical expressions lead to the scaling laws given by eqns. \eqref{Gp:scale} and \eqref{Gpp:scale}.
Although we have ignored the non-affine motion for crystalline solids, these analytical results quantitatively agree with those of the ordered MBS. 
Surprisingly, some characteristic features of disordered solids for low strain (or high pressure) can be captured.
These results indicate that the analysis of the toy model gives a basis for understanding the nonlinear rheology of frictional granular materials under small strain.

Although the values of the plateaus in $G'$ for disordered MBS depended on $\phi - \phi_{\rm J}$ \cite{OHern02,Otsuki17}, the corresponding values of the TBS are independent of $\phi - \phi_{\rm J}$, as expressed in eqn. \eqref{GpT}.
In addition, our analytical expressions cannot reproduce the second decrease of $G'$ and increase of $G''$ near $\gamma_0 = 10^{-2}$ in the MBS.
The discrepancy should result from the disorder because it leads to the $\phi$-dependence of $G'$ \cite{Wyart}. 
To include the disorder effect, we regarded $k_{\rm t}/k_{\rm n}$ as a fitting parameter.
In previous studies on models with small degrees of freedom, e.g., the coherent potential approximation \cite{Goldenfeld,Yonezawa, Feng}, the corresponding fitting parameters were self-consistently determined.
In future studies, we will discuss the self-consistent determination of the parameter for the TBS.

Some researchers are interested in contributions from higher harmonics characterizing the nonlinear response to oscillatory shear \cite{Hyun2011}, but the nonlinear viscoelastic moduli characterizing the higher harmonics are negligibly small for jammed frictionless particles \cite{Otsuki22}.
However, the higher harmonics in the frictional granular materials  require further careful investigation.


\section*{Conflicts of interest}
There are no conflicts to declare.


\section*{Appendix A: Details of Ordered MBS}

\begin{figure}[htbp]
\centering
\includegraphics[width=0.5\linewidth]{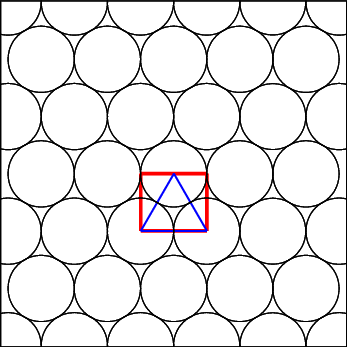}
  \caption{
    Initial configuration of mono-dispersed particles on a triangular lattice.
The red rectangle, including interactions represented by the blue lines, corresponds to the TBS.
}
  \label{confCR}
\end{figure}

This section explains the details of the ordered MBS consisting of monodispersed particles initially placed on a triangular lattice.
We consider a two-dimensional assembly of $N$ frictional particles in a periodic box with sizes along the $x$ and $y$ directions $L_x$ and $L_y$, respectively.
Here, we initially place $N=2 N_x N_y$ particles of diameter $d$ with integers $n_x$ and $n_y$ at $\boldsymbol r_i$ as
\begin{equation}
  \boldsymbol r_i = \left (n_x \ell - L_x/2, \sqrt{3} n_y \ell -L_y/2 \right )
\end{equation}
for $0 \le i < N_x N_y$ with integers $n_x$, $n_y$, and $i = n_x + N_x n_y$.
For $N_x N_y \le i < 2 N_x N_y$, $\boldsymbol r_i$ is defined as
\begin{equation}
  \boldsymbol r_i = \left ((n_x + 1/2)\ell - L_x/2, \sqrt{3} (n_y+ 1/2) \ell -L_y/2 \right )
\end{equation}
with $i = n_x + N_x n_y + N_x N_y$.
The initial configuration is illustrated in Fig. \ref{confCR}.
We choose $L_x = N_x \ell$ and $L_y = \sqrt{3} N_y \ell$ with $\ell=d(1-\epsilon)$.

The position $\boldsymbol r_i$ and peculiar momentum ${\boldsymbol p}_i$ of particle $i$ with mass $m_i$ and diameter $d_i$ are driven by the SLLOD equation under the Lees-Edwards boundary condition as \cite{Evans}
\begin{eqnarray}
\label{ri}
  \frac{d}{dt} {\boldsymbol r}_i & = & \dot \gamma(t) y_i \boldsymbol e_x + \frac{\boldsymbol p_i}{m_i}, \\
\label{pi}
   \frac{d}{dt} {\boldsymbol p}_i & = & - \dot \gamma(t) p_{i,y} \boldsymbol e_x + \boldsymbol f_i,
\end{eqnarray}
where $\dot \gamma(t) = \gamma_0 \omega \cos \omega t$ and $\bm{e}_x = (1,0)$ is the unit vector along the $x$ direction.
The interaction force $\boldsymbol f_i$ is defined as 
\begin{equation}
  \label{FMBS}
  \boldsymbol f_i = \sum_{j \neq i} \left ( f_{ij}^{\rm (n)} \boldsymbol n_{ij} + f_{ij}^{\rm (t)}\boldsymbol t_{ij}  \right ) H(d_{ij} - r_{ij})
\end{equation}
with $d_{ij} = (d_i + d_j)/2$, $\boldsymbol n_{ij} = \boldsymbol r_{ij} / r_{ij}$, $\boldsymbol t_{ij} = (-n_{ij,y},n_{ij,x})$, and $\boldsymbol r_{ij} = \boldsymbol r_i -  \boldsymbol r_j = (x_{ij},y_{ij})$.
The normal force is given by
\begin{equation}
 f_{ij}^{\rm (n)} = -\left ( k_{\rm n} u^{\rm (n)}_{ij}
 + \eta_{n} v^{\rm (n)}_{ij}\right )
  \label{FnMBS}
\end{equation}
with a normal viscous constant $\eta_{\rm n}$ and
\begin{eqnarray}
  v^{\rm (n)}_{ij} = \left ( \boldsymbol v_i - \boldsymbol v_j \right ) \cdot \boldsymbol n_{ij},
\end{eqnarray}
where the velocity of particle $i$ is given by $\boldsymbol v_i = \frac{d}{dt} \boldsymbol r_i$.
The following model is adopted for the tangential force:
\begin{equation}
  f_{ij}^{\rm (t)} = {\rm min} \left ( |\tilde f_{ij}^{\rm (t)}|, \mu f_{ij}^{\rm (n,el)} \right ) {\rm sgn} (\tilde f_{ij}^{\rm (t)}),
  \label{FtMBS}
\end{equation}
where $f_{ij}^{\rm (n,el)} = - k_{\rm n} u^{\rm (n)}_{ij}$ denotes the elastic part of the normal force.
Here, $\tilde f_{ij}^{\rm (t)}$ is given by
\begin{equation}
  \tilde f_{ij}^{\rm (t)} = - \left ( k_{\rm t} u_{ij}^{\rm (t)} + \eta_{\rm t} v_{ij}^{\rm (t)} \right )
\end{equation}
with a tangential viscous constant $\eta_{\rm t}$.
The tangential velocity $v_{ij}^{\rm (t)}$ is given by 
\begin{equation}
  v_{ij}^{\rm (t)} = (\boldsymbol v_i - \boldsymbol v_j)\cdot \boldsymbol t_{ij}.
  \label{VT}
\end{equation}
The tangential displacement $u_{ij}^{\rm (t)}$ satisfies $\frac{d}{dt} u_{ij}^{\rm (t)} = v_{ij}^{\rm (t)}$ for $|\tilde f_{ij}^{\rm (t)}|< \mu f_{ij}^{\rm (n,el)}$, whereas $u_{ij}^{\rm (t)}$ remains unchanged for $|\tilde f_{ij}^{\rm (t)}| \ge \mu f_{ij}^{\rm (n,el)}$.
The tangential displacement $u_{ij}^{\rm (t)}$ is set to zero if $i$ and $j$ are detached.

If all the particles are separated, the packing fraction $\phi$ for the ordered MBS is defined as
\begin{align}
\label{phi}
    \phi = \frac{\sum_i \pi d_i^2 }{4L_x L_y}.
\end{align}
Even if contacts exist between the particles, we use eqn. \eqref{phi} by assuming that the contact length $d_{ij} - r_{ij}$ is sufficiently lower than $d_{ij}$.
Using eqn. \eqref{phi}, $\phi$ is defined as
\begin{align}
    \phi = \frac{\pi}{2 \sqrt{3}(1-\epsilon)^2 }.
\end{align}
The jamming point of this system is 
\begin{align}
    \phi_{\rm J} = \frac{\pi}{2 \sqrt{3}}
\end{align}
with $\epsilon = 0$.
The distance from the jamming point is proportional to $\epsilon$.
\begin{align}
    \phi - \phi_{\rm J} \simeq \frac{\pi}{ \sqrt{3}} \epsilon
\end{align}
for $\epsilon \ll 1$.

The shear stress $\sigma$ is defined by eqn. \eqref{s} in the main article with
the normal component 
\begin{equation}
  \sigma^{\rm (n)} = - \frac{1}{ L_x L_y} \sum _i \sum_{j>i} \frac{x_{ij} y_{ij}}{r_{ij}} f^{\rm (n)}_{ij}
  \label{Sn:MBS}
\end{equation}
and tangential component
\begin{equation}
  \sigma^{\rm (t)} = - \frac{1}{ 2 L_x L_y} \sum _i \sum_{j>i} \frac{x_{ij}^2 -y_{ij}^2}{r_{ij}}f^{\rm (t)}_{ij}.
  \label{St:MBS}
\end{equation}
The pressure is defined as
\begin{equation}
  P = \frac{1}{2 L_x L_y} \sum _i \sum_{j>i} (x_{ij} f_{ij,x}+y_{ij} f_{ij,y}).
  \label{P:MBS}
\end{equation}

We use $N_x = 8$, $N_y = 4$, $N_{\rm c}=20$, $k_{\rm t} = k_{\rm n}$, and $\eta_{\rm n} =\eta_{\rm t} = k_{\rm n} \sqrt{m/k_{\rm n}}$, where $m$ denotes the mass of a particle with diameter $d$.
This model corresponds to a restitution coefficient $e = 0.043$.
We adopt the leapfrog algorithm considering a time step of $\Delta t = 0.05 t_0$.
We choose $ \omega = 1.0 \times 10^{-4} \sqrt{k_{\rm n}/m}$ as the quasistatic shear deformation because $G'$ and $G''$ are almost independent of $ \omega$ for $\omega \le 1.0 \times 10^{-3} \sqrt{k_{\rm n}/m}$.

As shown in Figs. \ref{Gp_3P} and \ref{Gpp_3P}, the behaviors of $G'$ and $G''$ of the TBS agree with that of the MBS.
We explain the theoretical background of the TBS.
The initial configuration is shown in Fig. \ref{confCR}; it contains the unit cell represented by the red rectangle with length $\ell$ and height $\sqrt{3} \ell /2$. 
It contains interactions between the three particles represented by blue lines. 
Here, we assume that the particles move affinely as
\begin{align}
   \boldsymbol r_{i}(t) = \boldsymbol r_{i}(0) + \gamma(\theta(t)) y_i(0) \boldsymbol e_x.
\end{align}
In this case, the corresponding relative distances between the particles in any unit cell are identical.

In particular, in a unit cell containing particles $i=i_1, i_2$, and $i_3$ with $i_1 = N_x N_y$, $i_2 = 0$, and $i_3 = 1$, the positions of the particles are given by
\begin{align}
\label{TEE1}
  & \boldsymbol r_{i_1}(t)  =  \left ( \gamma(\theta(t))\left ( \frac{\sqrt{3}\ell - L_y}{2} \right ) + \frac{\ell - L_x}{2}, \frac{\sqrt{3}\ell - L_y}{2}\right), \\
  & \boldsymbol r_{i_2}(t)  =  \left ( - \gamma(\theta(t)) \frac{L_y}{2} - \frac{L_x}{2},  - \frac{L_y}{2} \right), \\
\label{TEE3}
  & \boldsymbol r_{i_3}(t)  =  \left ( - \gamma(\theta(t)) \frac{L_y}{2} + \ell - \frac{L_x}{2},  -\frac{L_y}{2} \right).
\end{align}
The relative distances between these particles are identical to those of the TBS, given by eqns. \eqref{TE1}-\eqref{TE3}, which indicates that the TBS provides the interaction forces among the three particles.
This system includes $2 N_x N_y$ unit cells with identical interaction forces. 
Hence, the normal and tangential components of $\sigma$ are given by 
\begin{align}
  & & \sigma^{\rm (n)}  =  - \frac{ 2 N_x N_y}{ L_x L_y} \sum _{i = i_1, i_2, i_3} \left \{  \sum_{\substack{j=i_1, i_2, i_2 \\ (j>i)} } \frac{x_{ij}y_{ij}}{r_{ij}} f^{\rm (n)}_{ij} \right \}, \\
& & \sigma^{\rm (t)}  =  - \frac{ N_x N_y}{  L_x L_y} \sum _{i = i_1, i_2, i_3} \left \{ \sum_{\substack{j=i_1, i_2, i_2 \\ (j>i)} } \frac{x_{ij}^2-y_{ij}^2}{r_{ij}} f^{\rm (t)}_{ij}\right \}.
\end{align}
The pressure is also given by:
\begin{equation}
  P = \frac{N_x N_y}{ L_x L_y}  
  \sum _{i = i_1, i_2, i_3} \left \{  \sum_{\substack{j=i_1, i_2, i_2 \\ (j>i)} }
   (x_{ij} f_{ij,x}+y_{ij} f_{ij,y}) \right \}.
\end{equation}
Using the relation $L_x L_y / (2 N_x N_y) = \sqrt{3} \ell ^2 / 2$ corresponding to $A= \sqrt{3} \ell ^2 / 2$, $\sigma^{\rm (n)}$, $\sigma^{\rm (t)}$, and $P$ coincide with eqns. \eqref{s}-\eqref{P}.
Hence, if the assumptions of the affine motion, i.e., eqns. \eqref{TEE1}--\eqref{TEE3}, are satisfied, $G'$ and $G''$ in the ordered MBS coincide with those in the TBS.

\section*{Appendix B: Effect of particle rotation}

In this section, we illustrate the effect of particle rotation, which was not described in Appendix A.
In the model with rotation, the tangential velocity $v_{ij}^{\rm (t)}$ is given by 
\begin{equation}
  v_{ij}^{\rm (t)} = (\boldsymbol v_i - \boldsymbol v_j)\cdot \boldsymbol t_{ij} - (d_i \omega_i + d_j \omega_j)/2
\end{equation}
instead of eqn. \eqref{VT},
where $\omega_i$ denotes the angular velocity of particle $i$.
The time evolution of $\omega_i$ is given by
\begin{equation}
   I_i \frac{d}{dt} \omega_i = T_i 
   \label{omegai}
\end{equation}
with the moment of inertia $I_i = m_i d_i^2 /8$ and torque $T_i =  - \sum_j \frac{d_i}{2} \boldsymbol F_{ij}^{\rm (t)} \cdot \boldsymbol t_{ij}$.

\begin{figure}[htbp]
\centering
\includegraphics[width=0.7\linewidth]{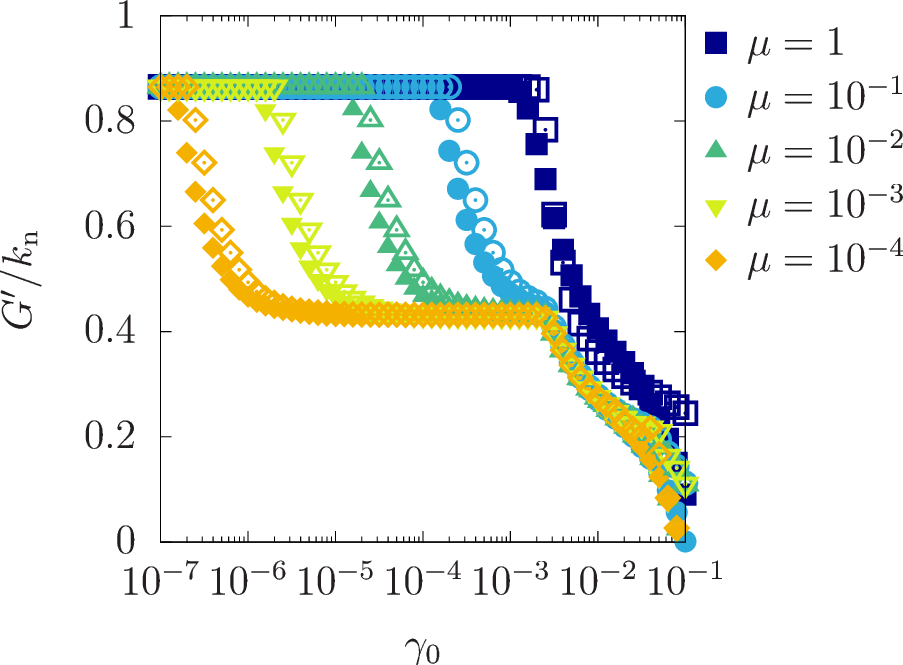}
  \caption{
    Storage modulus $G'$ against $\gamma_0$ for the ordered MBS with $\mu = 0.01$ and $\epsilon = 0.001$.
    The open and filled symbols represent the results of the particles with and without rotation, respectively.
}
  \label{Gp_CR}
\end{figure}

As shown in Fig. \ref{Gp_CR}, we plot $G'$ in the ordered MBS with and without rotation with $k_{\rm t}/k_{\rm n}=1.0$ and $\epsilon = 0.001$ for various values of $\mu = 0.01$.
The values of other parameters are the same as those in Appendix A.
The effect of particle rotation is negligible, except for the region near $\gamma_c$.

\begin{figure}[htbp]
\centering
\includegraphics[width=0.7\linewidth]{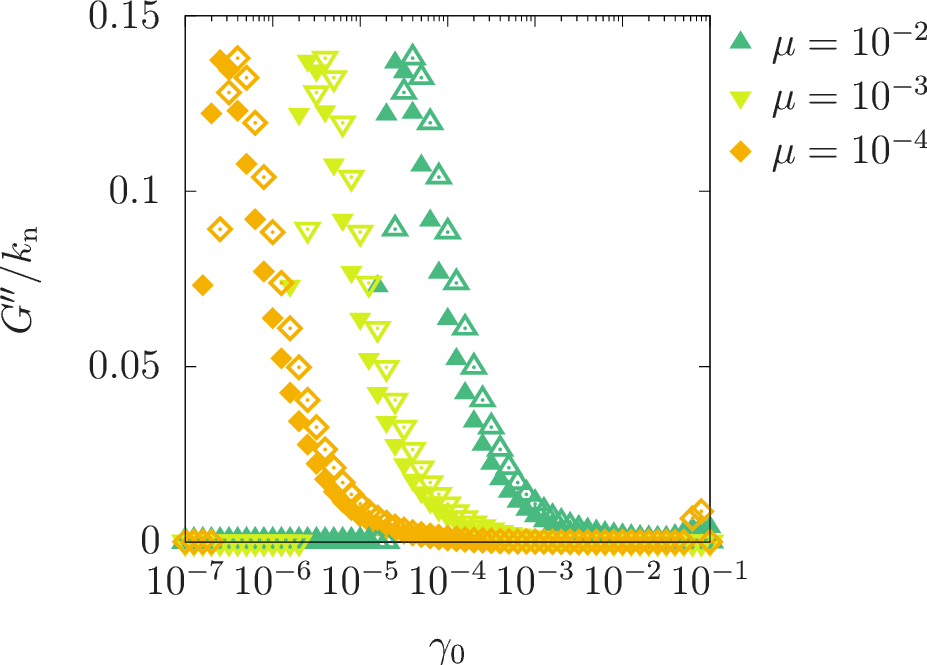}
  \caption{
    Loss modulus $G''$ against $\gamma_0$ for the ordered MBS with $\mu = 10^{-2},10^{-3},10^{-4},10^{-5}$ and $\epsilon = 0.001$.
    The open and filled symbols represent the results of the particles with and without rotation, respectively.
}
  \label{Gpp_CR1}
\end{figure}

As shown in Fig. \ref{Gpp_CR1}, we plot $G''$ in the ordered MBS with and without rotation with $k_{\rm t}/k_{\rm n}=1.0$ and $\epsilon = 0.001$ for $\mu = 10^{-2},10^{-3},10^{-4},10^{-5}$ and $0.00001$.
The values of other parameters are the same as those in Appendix A.
There are slight deviations in the peak position near $\gamma_c$ between particles with and without rotation.

\begin{figure}[htbp]
\centering
\includegraphics[width=0.7\linewidth]{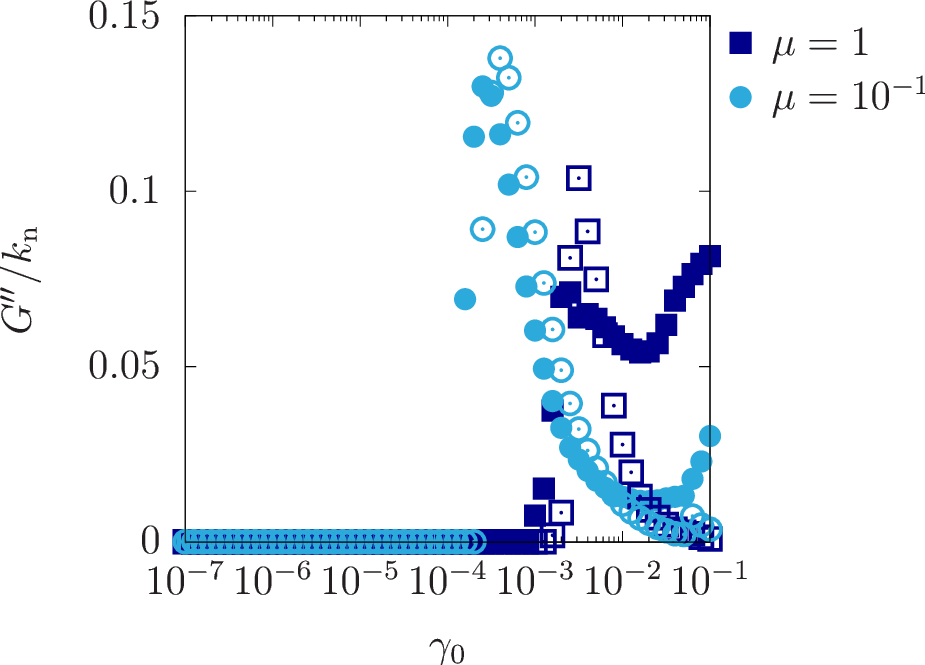}
  \caption{
    Loss modulus $G''$ against $\gamma_0$ for the ordered MBS with $\mu = 1.0, 0.1$ and $\epsilon = 0.001$.
    The open and filled symbols represent the results of the particles with and without rotation, respectively.
}
  \label{Gpp_CR2}
\end{figure}

In Fig. \ref{Gpp_CR2}, we plot $G''$ in the ordered MBS with and without rotation with $k_{\rm t}/k_{\rm n}=1.0$ and $\epsilon = 0.001$ for $\mu = 1.0$ and $0.1$.
There are slight deviations in the peak position near $\gamma_c$ between particles with and without rotation even for these higher $\mu$.
In addition, the second increase in $G''$ around $\gamma_0 = 0.1$ for particles without rotation disappears for those with rotation.

\section*{Appendix C: Analytical calculation of shear stress and pressure}

This section briefly explains the derivation of the normal and tangential components of shear stress and pressure for a small value of $\gamma_0$.
From eqns. \eqref{TE1}--\eqref{TE3}, the relative distance $\boldsymbol{r}_{ij}(t)$ is given by
\begin{align}
  \label{r12}
  & \boldsymbol{r}_{12}(\theta(t)) = \left ( \frac{\sqrt{3} \gamma(\theta(t)) + 1 }{2}\ell, \frac{\sqrt{3} \ell}{2} \right ), \\
  \label{r13}
  & \boldsymbol{r}_{13}(\theta(t)) = \left ( \frac{\sqrt{3} \gamma(\theta(t)) -1 }{2}\ell, \frac{\sqrt{3} \ell}{2} \right ), \\
  \label{r23}
  &   \boldsymbol{r}_{23}(\theta(t)) = \left ( -\ell, 0 \right ).
\end{align}
Substituting these equations into $u^{\rm (n)}_{ij} = r_{ij} - d$, the normal displacements are given by 
\begin{align}
  & u^{\rm (n)}_{12}(t) = - \epsilon d + \frac{\sqrt{3}}{4} \ell \gamma(\theta(t)) + O(\gamma_0^2), \\
      & u^{\rm (n)}_{13}(t) = - \epsilon d - \frac{\sqrt{3}}{4} \ell \gamma(\theta(t)) + O(\gamma_0^2), \\
  & u^{\rm (n)}_{23}(t) = - \epsilon d.
\end{align}
Substituting these equations into eqn. \eqref{Fn}, we obtain
the normal force as
\begin{align}
  \label{Fn0}
  & f_{12}^{\rm (n)} = k_{\rm n}  \left ( \epsilon d - \frac{\sqrt{3}}{4} \gamma(\theta) \ell \right ), \\
  \label{Fn1}
  & f_{13}^{\rm (n)} = k_{\rm n}  \left ( \epsilon d - \frac{\sqrt{3}}{4} \gamma(\theta) \ell \right ), \\
  \label{Fn2}
  & f_{23}^{\rm (n)} = k_{\rm n}  \epsilon d
\end{align}
up to $O(\gamma_0)$.

By differentiating eqns. \eqref{r12}--\eqref{r23} with time $t$, we obtain the relative velocity as
\begin{align}
  \label{v01}
  & \boldsymbol{v}_{12}(t) = \left ( \frac{\sqrt{3} \dot \gamma(\theta(t))\ell }{2}, 0 \right ), \\
  \label{v13}
  & \boldsymbol{v}_{13}(t) = \left ( \frac{\sqrt{3} \dot \gamma(\theta(t))\ell }{2}, 0 \right ), \\
  \label{v11}
  &   \boldsymbol{v}_{23}(t) = \left ( 0, 0 \right )
\end{align}
with the strain rate $\dot \gamma(\theta(t)) = \frac{d}{dt} \gamma(\theta(t))$.
The tangential unit vector is given by
\begin{align}
  \label{t01}
  & \boldsymbol{t}_{12}(t) = \left (-\frac{\sqrt{3} \ell}{2}, \frac{\sqrt{3} \gamma(\theta(t)) + 1 }{2}\ell \right )/ \left |\boldsymbol{r}_{12} \right |, \\
  \label{t13}
  & \boldsymbol{t}_{13}(t) = \left (-\frac{\sqrt{3} \ell}{2}, \frac{\sqrt{3} \gamma(\theta(t)) - 1 }{2}\ell \right )/ \left |\boldsymbol{r}_{13} \right |, \\
  \label{t11}
  &   \boldsymbol{t}_{23}(t) = \left ( 0, -1 \right ).
\end{align}
By considering the inner product of $\boldsymbol{v}_{ij}$ and $\boldsymbol{t}_{ij}$, the tangential velocity is given by
\begin{align}
  & v^{\rm (t)}_{12}(t) = - \frac{3}{4} \ell \dot \gamma(\theta(t)) + O(\gamma_0^2), \\
  & v^{\rm (t)}_{13}(t) = - \frac{3}{4} \ell \dot \gamma(\theta(t)) + O(\gamma_0^2), \\
  & v^{\rm (t)}_{23}(t) = 0.
\end{align}

If the transition from the stick state to the slip state does not occur under oscillatory shear, the tangential displacement is obtained by integrating $v^{\rm (t)}_{ij}(t)$ as
\begin{align}
  & u^{\rm (t)}_{12}(t) = u^{\rm (t)}_{13}(t) = - \frac{3}{4} \ell \gamma(\theta(t)) + O(\gamma_0^2), \\
  & u^{\rm (t)}_{23}(t) = 0.
\end{align}
Substituting these equations into $f_{ij}^{\rm (t)} = - k_{\rm t} u_{ij}^{\rm (t)}$ yields
\begin{align}
  \label{Ft0}
  & f_{12}^{\rm (t)} = f_{13}^{\rm (t)} =   3k_{\rm t} \gamma(\theta (t)) \ell/4, \\
  \label{Ft1}
  & f_{23}^{\rm (t)} =0
\end{align}
up to $O(\gamma_0)$.
The condition that the transition does not occur is satisfied when
$f_{12}^{\rm (t)} < \mu f_{12}^{\rm (n)}$ for $\gamma = \gamma_0$.
Using eqns. \eqref{Fn0} and \eqref{Ft0} with the assumption $\gamma_0 \ll \epsilon$, the condition is replaced by $\gamma_0 < \gamma_c$ with $\gamma_c$ given by eqn. \eqref{gc}.

For $\gamma_0 > \gamma_c$, there exist regions where $u_{ij}^{\rm (t)}$ is unchanged in the slip state as
\begin{align}
  \label{ut0}
  &u_{12}^{\rm (t)} =
   \left\{
\begin{array}{ll}
  -\dfrac{\mu k_{\rm n} \epsilon d}{k_{\rm t}}, & 0 \le \theta(\theta) < \dfrac{\pi}{2} \\
-\dfrac{\mu k_{\rm n} \epsilon d}{k_{\rm t}}
- \dfrac{3d (\gamma(\theta) - \gamma_0)}{4}, & \dfrac{\pi}{2} \le \theta < \dfrac{\pi}{2} + \Theta \\
   \dfrac{\mu k_{\rm n} \epsilon d}{k_{\rm t}}, & \dfrac{\pi}{2} + \Theta \le \theta <  \dfrac{3\pi}{2}\\
    \dfrac{\mu k_{\rm n} \epsilon d}{k_{\rm t}}
- \dfrac{3d (\gamma(\theta) + \gamma_0)}{4} , & \dfrac{3\pi}{2} \le \theta <  \dfrac{3\pi}{2} + \Theta\\
   -\dfrac{\mu k_{\rm n} \epsilon d}{k_{\rm t}}, & \dfrac{3\pi}{2} + \Theta \le \theta <  2\pi,
\end{array}
\right. \\
   & u_{13}^{\rm (t)} = u_{12}^{\rm (t)}, \\
  \label{ut1}
  &u_{23}^{\rm (t)} = 0,
\end{align}
where $\Theta$ satisfies
\begin{align}
-\frac{\mu k_{\rm n} \epsilon d}{k_{\rm t}}
- 3d\frac{ \gamma\left (\frac{\pi}{2} + \Theta \right) - \gamma_0}{4}
=
 \frac{\mu k_{\rm n} \epsilon d}{k_{\rm t}}.
\end{align}
This equation provides
$\Theta = \cos^{-1} \left ( 1 - 2 \gamma_c / \gamma_0 \right )$.
Substituting these equations into $f_{ij}^{\rm (t)} = - k_{\rm t} u_{ij}^{\rm (t)}$ yields
\begin{align}
  \label{Ftt0}
  &f_{12}^{\rm (t)} =
   \left\{
\begin{array}{ll}
  -\mu k_{\rm n} \epsilon d, & 0 \le \theta(\theta) < \dfrac{\pi}{2} \\
-\mu k_{\rm n} \epsilon d
- \dfrac{3k_{\rm t} d (\gamma(\theta) - \gamma_0)}{4}, & \dfrac{\pi}{2} \le \theta < \dfrac{\pi}{2} + \Theta \\
   \mu k_{\rm n} \epsilon d, & \dfrac{\pi}{2} + \Theta \le \theta <  \dfrac{3\pi}{2}\\
    \mu k_{\rm n} \epsilon d
- \dfrac{3k_{\rm t} d (\gamma(\theta) + \gamma_0)}{4}, & \dfrac{3\pi}{2}  \le \theta <  \dfrac{3\pi}{2} + \Theta\\
   -\mu k_{\rm n} \epsilon d, & \dfrac{3\pi}{2} + \Theta \le \theta <  2\pi,
\end{array}
\right. \\
  \label{Ftt0d}
 & f_{13}^{\rm (t)}= f_{12}^{\rm (t)}, \\
  \label{Ftt1}
  &f_{23}^{\rm (t)} = 0.
\end{align}

The normal component of $\sigma$ in eqn. \eqref{Sn} is given by
\begin{equation}
    \sigma^{\rm (n)} = \sigma^{\rm (n)}_{12} + \sigma^{\rm (n)}_{13}
    \label{sn123}
\end{equation}
with
\begin{align}
  \label{sn12}
    & \sigma^{\rm (n)}_{12} = - \frac{1}{A} \frac{x_{12} y_{12}}{r_{12}} f_{12}^{\rm (n)} \\
  \label{sn13}
    & \sigma^{\rm (n)}_{13} = - \frac{1}{A} \frac{x_{13} y_{13}}{r_{13}} f_{12}^{\rm (n)}.
\end{align}
Substituting eqns. \eqref{r12} and \eqref{r13} with eqns. \eqref{Fn0} and \eqref{Fn1} into eqns. \eqref{sn12} and \eqref{sn13} and using eqn. \eqref{sn123}, we obtain $\sigma^{\rm (n)}$ as eqn. \eqref{sn}.

The tangential component of $\sigma$ in eqn. \eqref{St} is given by
\begin{equation}
    \sigma^{\rm (t)} = \sigma^{\rm (t)}_{12} + \sigma^{\rm (t)}_{13}
    \label{st123}
\end{equation}
with
\begin{align}
  \label{st12}
    & \sigma^{\rm (t)}_{(12)} = - \frac{1}{2 A} \frac{x_{12}^2 -  y_{12}^2}{r_{12}} f_{12}^{\rm (t)} \\
  \label{st13}
    & \sigma^{\rm (t)}_{(13)} = - \frac{1}{2 A} \frac{x_{13}^2 - y_{13}^2}{r_{13}} f_{12}^{\rm (t)}.
\end{align}
Substituting eqns. \eqref{r12} and \eqref{r13} with eqn. \eqref{Ft0} into eqns. \eqref{st123}, \eqref{st12}, and \eqref{st13}, we obtain $\sigma^{\rm (t)}$ as eqn. \eqref{st1} for $\gamma_0 < \gamma_c$.
Using eqns. \eqref{Ftt0} and \eqref{Ftt0d} instead of eqn. \eqref{Ft0},
we obtain $\sigma^{\rm (t)}$ as eqn. \eqref{st2} for $\gamma_0 \ge \gamma_c$.

The pressure, i.e., $P$, in eqn. \eqref{P} is defined as
\begin{equation}
  P = P_{12} + P_{13} + P_{23}
  \label{P123}
\end{equation}
with
\begin{align}
  \label{Pij}
     P_{ij} =  \frac{1}{2 A} r_{ij} f_{ij}^{\rm (n)}.
\end{align}
Substituting eqns. \eqref{r12}-\eqref{r23} with eqns. \eqref{Fn0}-\eqref{Fn2} into eqns. \eqref{P123} and \eqref{Pij} with $\gamma = 0$, we obtain $P_0(\gamma_0,\mu)$ as eqn. \eqref{PT}.

\section*{Appendix D: Relation between shear modulus and stress--strain curve}

In this section, we relate the shape of the stress--strain curve to the complex shear modulus.
The shear stress $\sigma(\theta)$ is expanded using the Fourier series as
\begin{align}
  \sigma(\theta)  =  \gamma_0 \sum_{n=1}^{\infty}  G'_{n} \sin (n \theta)  + \gamma_0 \sum_{n=1}^{\infty} G''_{n} \cos(n \theta),
\end{align}
where $G'_{n}$ and $G''_n$ with $n>1$ denote the higher harmonics, $G' = G'_1$, and $G'' = G''_1$.
By neglecting $G'_{n}$ and $G''_n$ for $n>1$,
\begin{align}
  G' \simeq \frac{\sigma\left (\theta=\pi/2 \right )}{\gamma_0} = \left . \frac{\sigma}{\gamma_0} \right |_{\gamma/\gamma_0 = 1} = \tilde \sigma_{\rm max},
\end{align}
which is the maximum value of the scaled stress--strain curve illustrated in Fig. \ref{st:Fig}(b).
This expression and the scaled stress--strain curve in Fig. \ref{st:Fig}(b) explain the decrease of $G'$ defined by eqn. \eqref{GpT}.

The area $S$ of the curve for $\sigma(\theta) / \gamma_0$ against $\gamma(\theta) / \gamma_0$ is given by
\begin{align}
  \label{S}
  S =  \int_{0}^{2 \pi} d \theta \frac{1}{\gamma_0} \frac{d \gamma(\theta)}{d \theta} \frac{\sigma (\theta)}{\gamma_0}.
\end{align}
Substituting eqn. \eqref{g} into eqn. \eqref{S} with eqn. \eqref{Gp}, we obtain	
\begin{align}
  S =  \int_{0}^{2 \pi} d \theta  \sigma (\theta) \cos \theta /\gamma_0 = \pi G'',
\end{align}
which results in $G'' = S / \pi$.
As $\gamma_0$ increases, the area $S$ of the scaled stress--strain curve in Fig. \ref{st:Fig}(b) increases first and decreases later, which explains the $\gamma_0$-dependence of $G''$ provided by eqn. \eqref{GppT}.

\section*{Appendix E: Details of Disordered MBS}

In this section, we present the details of the disordered MBS.
This model is an extension of the monodisperse model used in Appendix A, including the dispersion of the particles and disordered initial configuration.

The system is bidisperse and includes an equal number of particles with diameters $d$ and $d/1.4$.
To simulate the disordered MBS, we randomly place the particles in a rectangular box with an initial packing fraction of $\phi_{\rm I} = 0.75$.
The system is slowly compressed until the packing fraction reaches $\phi$ \cite{Otsuki21}.
In each compression step, the packing fraction is increased by $\Delta \phi = 1.0 \times 10^{-4}$ with an affine transformation. 
Thereafter, the particles are relaxed to a mechanical equilibrium state with the kinetic temperature $T_{\rm K} = \sum_i p_i^2 /(mN) < T_{\rm th}$. 
Here, we choose $T_{\rm th} = 1.0 \times 10^{-8}k_{\rm n}d^2$.
After compression, the oscillatory shear strain given by eqn. \eqref{g} is applied for $N_{\rm c}$ cycles.
In the last cycle, we measure $G'$ and $G''$ using eqns. \eqref{Gp} and \eqref{Gpp} with eqns. \eqref{s}--\eqref{St}.
The pressure, $P_0(\gamma_0,\mu)$ is obtained using eqn. \eqref{P} after the last cycle.
We use $\phi=0.87$, $N=1000$, $N_{\rm c}=20$, $L_y/L_x = 1$, $k_{\rm t} = 0.2k_{\rm n}$, and $\eta_{\rm n} =\eta_{\rm t} = k_{\rm n} \sqrt{m/k_{\rm n}}$.

\begin{figure}[htbp]
\centering
\includegraphics[width=0.7\linewidth]{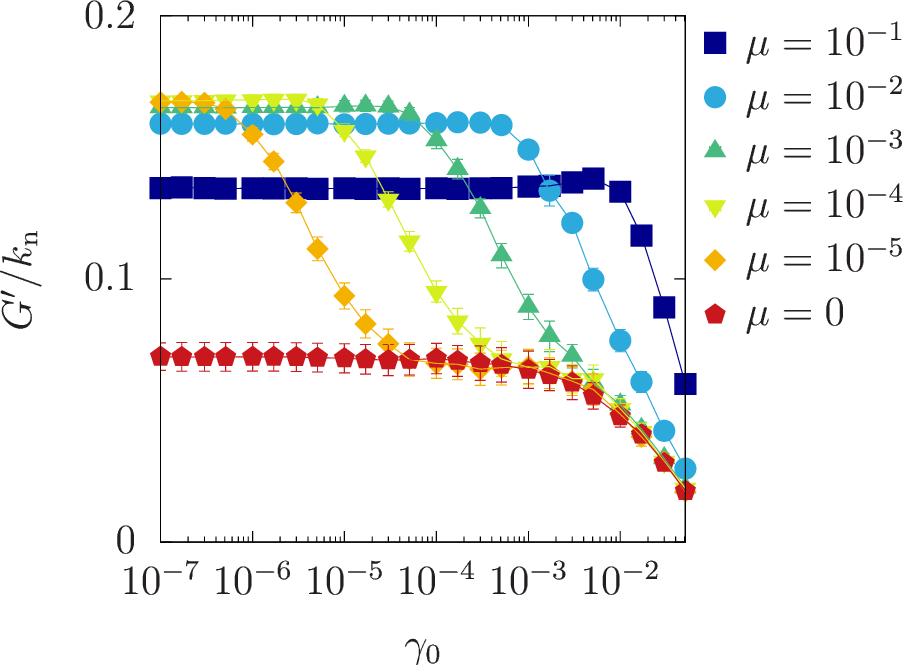}
  \caption{
    Storage modulus $G'$ in the disordered MBS against $\gamma_0$ with $\phi=0.870$ for various values of $\mu$.
}
  \label{Gp_YPT10}
\end{figure}

Figure \ref{Gp_YPT10} shows the storage modulus $G'$ against $\gamma_0$ in the disordered MBS for various values of $\mu$.
The storage modulus $G'$ is almost independent of $\gamma_0$ for a small $\gamma_0$ and decreases as $\gamma_0$ increases.
The endpoint of the first plateau increases with $\mu$ except for $\mu=0$.  
A second plateau of $G'$ exists for $\mu=10^{-4}$ and $10^{-5}$.
The behavior of $G'$ for relatively small $\gamma_0$ is similar to that of crystalline solids as depicted in Fig. \ref{Gp_3P}. On the other hand, the decrease of $G'$ for larger $\gamma_0$ cannot be captured by the analytical results of the TBS.
Note that $G'$ for $\mu=0.1$ in the limit $\gamma_0 \to 0$ is different from that for $\mu \le 0.01$, which results from the $\mu$ dependence of the jamming point $\phi_{\rm J}$ \cite{Otsuki17}.

\begin{figure}[htbp]
\centering
\includegraphics[width=0.7\linewidth]{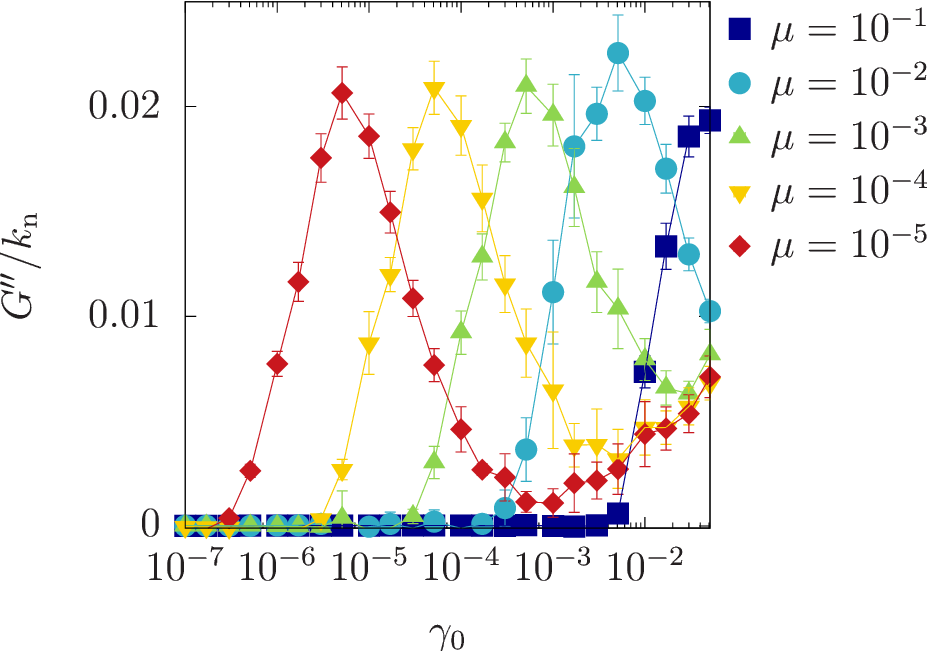}
  \caption{
    Loss modulus $G''$ in the disordered MBS against $\gamma_0$ with $\phi=0.870$ for various values of $\mu$.
}
  \label{Gpp_YPT10}
\end{figure}

Figure \ref{Gpp_YPT10} shows the loss modulus $G''$ in the disordered MBS against $\gamma_0$ for various values of $\mu$.
For sufficiently small $\gamma_0$, $G''$ is zero, while
$G''$ becomes non-zero as $\gamma_0$ increases. 
The loss modulus $G''$ starts to increase for smaller $\gamma_0$ as $\mu$ decreases.
Similar to the case of $G'$, TBS captures only the behavior of relatively small $\gamma_0$ (see Figs. \ref{Gpp_3P} and \ref{Gpp_YPT10}).

\section*{Appendix F: Numerical shear modulus for TBS}

In this section, we show the behaviors of $G'$ and $G''$ in the TBS without the assumption used to obtain the analytical solution.
Here, we numerically obtain $G'$ and $G''$ under quasistatic oscillation using eqns. \eqref{Gp} and \eqref{Gpp} based on the left Riemann sum, where the integration of $\Psi(\theta)$, i.e.,
\begin{align}
    \int_0^{2 \pi} d \theta \ \Psi (\theta),
\end{align}
is approximated as
\begin{align}
    \int_0^{2 \pi} d \theta \ \Psi (\theta)
   \simeq \sum_{n=1}^M \Psi(\theta_n) \Delta \theta 
\end{align}
with $\Delta \theta = 2 \pi / M$ and $\theta_n =  (n-1)\Delta \theta$.
We use $\epsilon = 0.001$ and $ \Delta \theta = 5.0 \times 10^{-5}$ in our simulation.

\begin{figure}[htbp]
\centering
\includegraphics[width=0.7\linewidth]{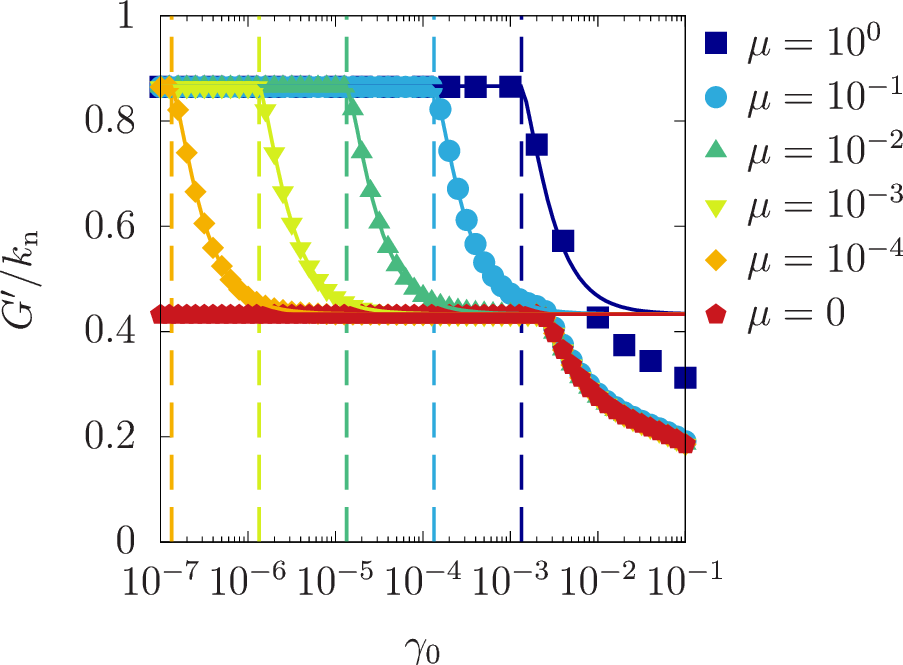}
  \caption{
    Storage modulus $G'$ against $\gamma_0$ with $k_{\rm t} / k_{\rm n} = 1.0$ and $\epsilon = 0.001$ for various values of $\mu$.
    The points represent the numerical results of the TBS, while the thin solid lines represent the analytical result given by eqn. \eqref{GpT}. 
    The vertical dashed lines represent the critical amplitude $\gamma_c(\mu)$ given by eqn. \eqref{gc} for $\mu = 10^{-4}, 10^{-3}, 10^{-2}, 10^{-1}, 10^{0}$ from left to right.
}
  \label{Gp_3PD}
\end{figure}

As shown in Fig. \ref{Gp_3PD}, we plot the storage modulus $G'$ numerically obtained from the TBS against $\gamma_0$ with $k_{\rm t} / k_{\rm n} = 1.0$ for various values of $\mu$ as points.
Moreover, we plot the analytical results derived from eqn. \eqref{GpT} as thin solid lines.
The numerical results agree with the analytical results for $\gamma_0 < 0.003$ and reproduce the second plateau of the MBS shown in Fig. \ref{Gp_3P}.

\begin{figure}[htbp]
\centering
\includegraphics[width=0.7\linewidth]{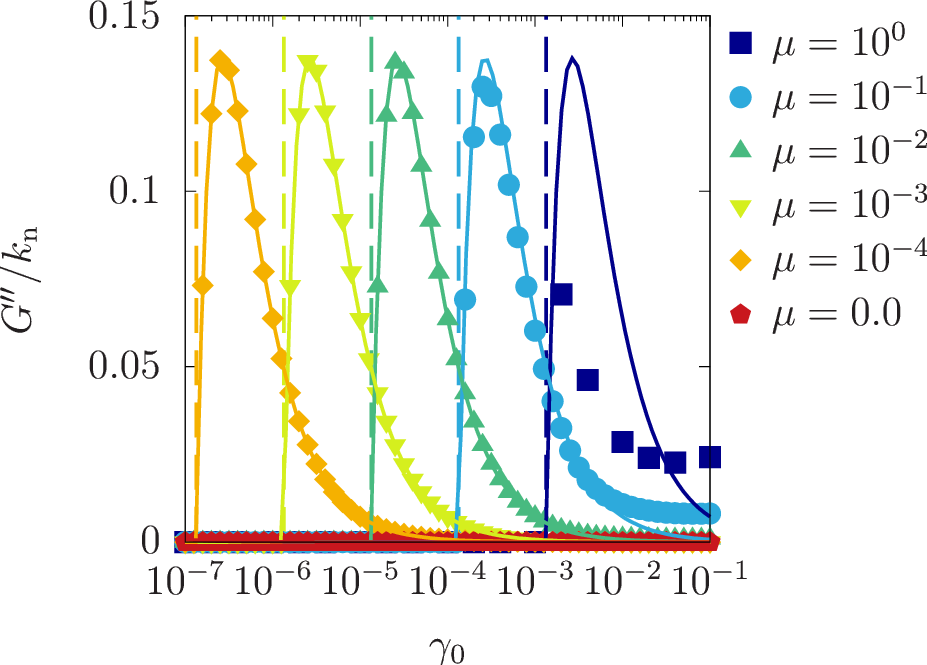}
  \caption{
    Loss modulus $G''$ against $\gamma_0$ with $k_{\rm t} / k_{\rm n} = 1.0$ and $\epsilon = 0.001$ for various values of $\mu$.
    The points represent the numerical results of the TBS, while the thin solid lines represent the analytical results obtained from eqn. \eqref{GppT}. 
    The vertical dashed lines represent the critical amplitude $\gamma_c(\mu)$ obtained from eqn. \eqref{gc} for $\mu = 10^{-4}, 10^{-3}, 10^{-2}, 10^{-1}, 10^{0}$ from left to right.
}
  \label{Gpp_3PD}
\end{figure}

Figure \ref{Gpp_3PD} shows the loss modulus $G''$ numerically obtained from the TBS against $\gamma_0$ with $k_{\rm t} / k_{\rm n} = 1.0$ for various values of $\mu$ as points.
We also plot the analytical results given by eqn. \eqref{GppT} as thin solid lines.
The numerical results agree with the analytical results for $\gamma_0 < 0.003$.

\section*{Acknowledgements}
The authors thank K. Saitoh, D. Ishima, and S. Takada for fruitful discussions.
This study was supported by JSPS KAKENHI under Grant Nos. JP19K03670 and JP21H01006.



\balance


\bibliography{export} 
\bibliographystyle{rsc} 

\end{document}